\documentclass{article}
\usepackage{jcappub}
\usepackage{graphicx}
\usepackage{amsmath}

\newcommand{\be}{\begin{equation}}
\newcommand{\ee}{\end{equation}}
\newcommand{\bea}{\begin{eqnarray}}
\newcommand{\eea}{\end{eqnarray}}
\newcommand{\h}{\mathcal{H}}

\title{Dark energy and dark matter perturbations in singular universes}

\author[a]{Tomasz Denkiewicz}
\emailAdd{atomekd@wmf.univ.szczecin.pl}
\affiliation[a]{\it Institute of Physics, University of Szczecin, Wielkopolska 15,
          70-451 Szczecin, Poland}
\affiliation[a]{\it Copernicus Center for Interdisciplinary Studies,
S{\l }awkowska 17, 31-016 Krak\'ow, Poland}
\date{\today}
\input epsf
\abstract{
  We discuss the evolution of density perturbations of dark matter and dark energy in cosmological models which admit future singularities in a finite time. Up to now geometrical tests of the evolution of the universe do not differentiate between singular universes and  $\Lambda$CDM scenario. We solve perturbation equations using the gauge invariant formalism. The analysis shows that the detailed reconstruction of the evolution of perturbations within singular cosmologies, in the dark sector, can exhibit important differences between the singular universes models and the $\Lambda$CDM cosmology. This is encouraging for further examination and gives hope for discriminating between those models with future galaxy weak lensing experiments like the Dark Energy Survey (DES) and Euclid or CMB observations like  PRISM and CoRE. 
}

\begin{document}
\maketitle

\section{Introduction}
\setcounter{equation}{0}
After the release of the supernova type Ia (SNIa) data in 1997 \cite{1998AJ,1999ApJ} number of works, which begin with the description of the dark energy problem, as recognized as, the one of the recent greatest problems of physics, started to grow. The answer to the question, what is the source of the recent accelerated expansion of the universe, confirmed later by other observations such as Cosmic Microwave Background \cite{cmb} and large scale structure \cite{lss}, is still pending. On the other hand, even before, already in 1986 a new type of singular behaviour of the universe, different than already known big bang/big crunch, was proposed \cite{BGT}. The idea gained more attention after 1997 and especially in recent years. Considering scenarios of the evolution of the universe, which is expanding in an accelerated manner, resulted with the discovery of new types of singularities like: a big-rip (BR), a sudden future singularity (SFS) \cite{barrow04, barrow042, gr-qc/0410033v3}, a generalized sudden future singularity (GSFS), a finite scale factor singularity (FSF) \cite{0910.0023v1} or a big-separation singularity (BS) \cite{not2005} and $w$-singularities \cite{0902.3107v3}. More recently, the Little Rip or Pseudo-Rip universes \cite{1106.4996v1,1112.2964v2}, where the destruction of all bound structures takes place, but dark energy density asymptotically reaches finite, value were found. Some other variations of those like grand rip and grand bang/crunch were studied too \cite{PhysRevD.90.064014}. Recently, it was also found that quintom  dark energy can drive quasi-rip singularity \cite{quasirip}. In this scenario disintegrated structures have the possibility to recombine after all.\\
Lately, scenarios with the future singularities were tested against the observations \cite{AO, DHD, GHDD, DDGH, FSF, laszlo2014, bamba, PhysRevD.79.083504, 2014PhRvD..90f3508L} like baryon acoustic oscillations (BAO) \cite{Eisenstein}, supernova type Ia or shift parameter \cite{shift}. For some of them, there were also done forecasts for the future data, like the redshift drift \cite{rd}, planned to be measured by ELT-HIRES (High-Resolution Ultra-stable Spectrograph for the E-ELT \cite{eelt}). The conclusion, that was drown was that the redshift drift test can play the role of a differential test between these models and $\Lambda$CDM, but still they are good candidates for the dark energy. The effects of running gravitational coupling and the entropic force on future singularities were considered in \cite{mirza}.  A possibility to vary the fine structure $\alpha$ within these models was also checked \cite{alpha}. Also the opportunity to avoid singularities due to quantum effects was investigated (see for example \cite{kiefer, PhysRevD.89.064016,PhysRevD.83.064027}) and possibilities of passing through the singularities were checked \cite{singularitycrossing,sc2,singhsing,PhysRevD.86.063522,MPDPLB11}. It seems that one of the less investigated topic is the evolution of density perturbations of the matter and the dark energy. Some of the linear density perturbations as the growth factor were tested, for FSFS scenario in \cite{1202.3280v2} and also for Big Rip, SFS, FSFS and Pseudo-Rip, in \cite{AO}. Nevertheless, it seems that, since singular behavior of the universe can mimic $\Lambda$CDM model regarding geometrical tests, the question of the evolution of perturbations is worth exploring in the purpose of finding further dynamical constraints onto those models.\\
Furthermore, it is encouraging that we are now in the so called epoch of precision cosmology and there are planned galaxy week lensing experiments like the Dark Energy Survey (DES) \cite{des} and Euclid \cite{euclid}. Those potentially will be able to discriminate between cosmological constant $\Lambda$CDM and evolving dark energy scenarios. Also, there are planned experiments to further explore cosmic background radiation in microwave to far-infrared bands in polarization and amplitude as Polarized Radiation and Imaging Spectroscopy (PRISM) \cite{prism} as well very high precision measurements of the polarization of the microwave sky are planned by Cosmic Origins Explorer (CoRE) satelite \cite{core}. All this measurements will improve constraints onto the dark sector. Surely, their constraining power gives some motivation to explore the perturbations in the dark sector. Having the results of such investigations will allow to benefit fully from aforementioned, forthcoming observations. 

In the present work, we explore the evolution of both dark energy and dark matter perturbations within SFS and FSF scenarios. Following \cite{Bardeen, Mukhanov, Starobinski} we use the gauge invariant formalism, for the evolution of the perturbations, introducing it in Section \ref{perturbations}. We work on the singular cosmologies and employ the parametrization of the scale factor introduced in \cite{barrow042}. Furthermore we work within the Friedmanian framework of isotropic cosmology. We assume that after the decoupling, to a good approximation the universe is filled with two barotropic fluids which behave as dark matter and dark energy. We introduce our models in Section \ref{themodels}. In Section \ref{perturbations evo} we investigate the evolution of perturbations in dark matter and energy. We plot the evolution of dark matter and dark energy perturbations for selected examples of FSFS and SFS cosmologies which are in the agreement with other observational tests like BAO, R, SNIa, Hubble parameter evolution. Chosen examples exhibit also the proper evolution of the adiabatic speed of sound and the barotropic equation of state parameter. Finally we give our conclusions in Section \ref{konk}.

\section{Gauge invariant perturbations}\label{perturbations}
\setcounter{equation}{0}
The cosmological perturbations, represented by the density contrast $\delta(k,z)=\delta\rho/ \rho$, are functions of the scale and the redshift. As such, they depend on the expansion rate and in principle the law of gravity on large scales. This is the reason for which the density evolution is a good tool to diagnose dynamical properties of the universe. It is encouraging that near future observations should highly improve constraints given on the dark matter density contrast, $\delta_m(k,z)$ \cite{deltaobservations}.\\
The cosmological perturbations, beyond the frame of Newtonian gravity and hydrodynamics were first studied by Lifshitz in 1946 \cite{Lifshitz:1945du}. Working with perturbations within the frame of general relativity brings the issue of the gauge freedom. In the work of Lifshitz the synchronous gauge was elaborated, which afterwards gained popularity. Later it occured, that working within this gauge could lead to spurious gauge modes or singularities. For that reason Bardeen developed the gauge invariant formalism in 1980 \cite{Bardeen}.
There are two most exploited types of gauge among the literature: the Newtonian called, also as the longitudinal gauge, and the synchronous gauge. Choice of the gauge is determined by the behaviour of inspected quantities, which in one gauge can have singular behavior and in another can behave in a fully regular way. 
In this work we will deploy gauge invariant formalism, which was used by many authors, following Bardeen \cite{Bardeen}.  Gauge invariant quantities are derived by constructing them as combinations of metric and energy momentum tensor. We will follow one of the popular notations \cite{Mukhanov}, ($4\pi G=c=1$):
\begin{equation}\label{Mukhanov1}
 \left.
\begin{array}{l}
\Delta\Psi - 3\mathcal{H}\left(\mathcal{H}\Phi + \Psi'\right) +
3\mathcal{K}\Psi = a^{2}\delta\rho\\
\mathcal{H}\Phi + \Psi' = a\left(\rho +
p\right)V\\
\left[\Psi'' + \mathcal{H}\Phi' + \left(2\mathcal{H}' +
\mathcal{H}^{2}\right)\Phi + 2\mathcal{H}\Psi' - \mathcal{K}\Psi +
\frac{1}{2}\Delta\left(\Phi - \Psi\right)\right]\delta^{i}_{j} - 
- \frac{1}{2}\gamma^{ik}\left(\Phi - \Psi\right)_{|kj}\\ = a^{2}\delta
p\delta^{i}_{j} - \sigma^{|i}_{|j},
\end{array}
\right.
\end{equation}
Here $\Phi(x,\eta)$ and $\Psi(x,\eta)$ are gauge invariant Bardeen potentials, which with the assumption that non diagonal, space-space components of the energy momentum tensor are absent, $\sigma=0$, are equal ie. $\Phi=\Psi$ and $a(\eta)$ is the scale factor as a function of conformal time $d\eta=da/a$; $\mathcal{H}(\eta)= a ^{\prime} / a$, with prime denoting derivative with respect to $\eta$; $\delta p$ and $\delta \rho$ are the perturbations of the pressure and energy density; $V(x,\eta)$ is the scalar potential of the velocity field; $\sigma(x,\eta)$ stands for the shear; $\gamma_{ij}$ stands for the spatial part of the metric and vertical bar denotes covariant derivative with respect to $\gamma_{ij}$; $\mathcal{K}$ is the curvature which is flat for $\mathcal{K}=0$, close for $\mathcal{K}=+1$ or open for $\mathcal{K}=-1$. Further we assume spatial flatness and adopt $\mathcal{K}=0$ case. We consider the linear regime of perturbations, that allow us to work in the Fourier space and track the evolution of each of the components independently.  

Aforementioned assumptions, together with the change of the independent variable, $\eta$ for $a$, allow us to transform set of equations, (\ref{Mukhanov1}), into the following set of equations for a multi-fluid system:
\begin{equation}\label{Muksys2}
\left.
\begin{array}{l}
-k^{2}\Phi - 3a\mathcal{H}^{2}\dot{\Phi} - 3\mathcal{H}^{2}\Phi = a^{2}
\sum_{i=1}^{N}\rho_{i}\delta_{i}\\
\mathcal{H}\Phi + a\h\dot{\Phi} = a\sum_{i =
1}^{N}\left(\rho_{i} + p_{i}\right)V_{i}\\
\left(a\mathcal{H}\right)^{2}\ddot{\Phi} + \left(4a\mathcal{H}^{2} +
a^{2}\h\dot{\h}\right)\dot{\Phi} + \left(2a\h\dot{\h} +
\mathcal{H}^{2}\right)\Phi = a^{2}\sum_{i=1}^{N}c_{{\rm
s}i}^{2}\rho_{i}\delta_{i},
\end{array}
\right.
\end{equation}
where the dot stands for the derivative with respect to the scale factor $a$ and $\delta_{i}$ is the density contrast defined in the following way:
\be\label{rhocont}
\delta_{i} \equiv \frac{\delta\rho_{i}}{\rho_{i}},
\ee
where  $\rho_{i}$ is the background energy density of the $i$th component.

In general, one can consider $N$ noninteracting fluids, each component is designated by its index $i$. Background pressure and density obey:

\be
\rho=\sum_{i=1}^N \rho_i, \quad p=\sum_{i=1}^N p_i
\ee
and for the perturbations we have:
\be\label{dpdrho}
\delta\rho = \sum_{i = 1}^{N}\delta\rho_{i},\quad  \delta p = \sum_{i =
1}^{N}\delta p_{i}.
\ee

Since we consider adiabatic perturbations of perfect, barotropic fluids we employ the following equation of state:
\be\label{steq}
\delta p_{i} = c_{{\rm s}i}^{2} \delta\rho_{i},
\ee
with the speed of sound given by: 
\be\label{cs}
c_{{\rm s}i}^{2} \equiv \frac{{\rm \partial}p_{i}}{{\rm
\partial}\rho_{i}}.
\ee
The speed of sound here is given at the constant entropy for the generic fluid component, $i$. \\
We will consider two fluid model, which consists of dark matter and dark energy. For $N>1$ the system (\ref{Muksys2}) is not fully determined and one needs additional equations. We assume that components of the content of the universe do not interact directly, that allows to close the system (\ref{Muksys2}) with the energy conservation equations and the Euler equations for each component. The continuity equation,
$\delta
T^{\mu}_{0 ;\mu} = 0
$, reads:
\begin{equation}\label{conteq}
\dot{\delta}_{i} + \frac{3}{a}\left(c_{{\rm s}i}^{2} -
w_{i}\right)\delta_{i} - 3\dot{\Phi}\left(1 + w_{i}\right) +
\frac{k^{2}}{a^{2}\h}\left(1 + w_{i}\right)V_{i} = 0
\end{equation}
and the Euler equation $\delta T^{\mu}_{l;\mu} = 0$, reads as follows:
\begin{equation}\label{Eulereq}
\left[\left(\rho_{i} + p_{i}\right)V_{i}\right]\dot{} +
\frac{3}{a}\left(\rho_{i} + p_{i}\right)V_{i} - \frac{c_{{\rm
s}i}^{2}\rho_{i}}{\h}\delta_{i} - \frac{\left(\rho_{i} +
p_{i}\right)}{\h}\Phi = 0,
\end{equation}
where $w_{i} \equiv p_{i}/\rho_{i}$.

Combining equations (\ref{Muksys2}), (\ref{conteq}) and (\ref{Eulereq}) it is possible to derive system of $N$ second order equations for the density contrast of each component $\delta_i$ cf. \cite{SolovStar,Starobinski}. 
 
In the case of $N = 2$ the resulting system has the following form:
\be\label{gensys}
\left.
\begin{array}{l}
\ddot{\delta}_{\rm 1} + \left(\frac{\dot{\h}}{\h} + \frac{A_{\rm 1}}{\h}\right)\dot{\delta}_{\rm 1} + \frac{B_{\rm 1}}{\h}\dot{\delta}_{\rm 2} + \frac{C_{\rm 1}}{\h^{2}}\delta_{\rm 1} + \frac{D_{\rm 1}}{\h^{2}}\delta_{\rm 2} = 0\\ \\
\ddot{\delta}_{\rm 2} + \left(\frac{\dot{\h}}{\h} + \frac{A_{\rm 2}}{\h}\right)\dot{\delta}_{\rm 2} + \frac{B_{\rm 2}}{\h}\dot{\delta}_{\rm 1} + \frac{C_{\rm 2}}{\h^{2}}\delta_{\rm 2} + \frac{D_{\rm 2}}{\h^{2}}\delta_{\rm 1} = 0,
\end{array}
\right.
\ee
where the coefficients $A_{\rm i},\ B_{\rm i},\ C_{\rm i},\ D_{\rm i}$ are given in the Appendix \ref{app}. As shown in \cite{SolovStar}, at the matter--dominated regime, the above coefficients are far
simpler and system (\ref{gensys}) can be rewritten in the following  quasi--Newtonian form:
\be\label{simp}
\left.
\begin{array}{l}
\ddot{\delta}_{\rm 1} + \left(\frac{\dot{\h}}{\h} +
\frac{2}{a}\right)\dot{\delta}_{\rm 1} +
\frac{k^{2}}{a^{2}\h^{2}}c_{\rm s1}^{2}\delta_{\rm 1} =
\frac{1}{\h^{2}}\left(\rho_{\rm 1}\delta_{\rm 1} +
\rho_{\rm 2}\delta_{\rm 2}\right)\\ \\
\ddot{\delta}_{\rm 2} + \left(\frac{\dot{\h}}{\h} +
\frac{2}{a}\right)\dot{\delta}_{\rm 2} +
\frac{k^{2}}{a^{2}\h^{2}}c_{\rm s2}^{2}\delta_{\rm 2} =
\frac{1}{\h^{2}}\left(\rho_{\rm 1}\delta_{\rm 1} + \rho_{\rm
2}\delta_{\rm 2}\right).
\end{array}
\right.
\ee
The other path to recover the  system (\ref{simp}) is to impose the assumption of the limit $k \gg \h$ \cite{SolovStar}. Although it is not necessary assumption being in the matter--dominated limit when following relations hold true: $p_{\rm 1} \ll \rho_{\rm 1}$, $p_{\rm 2} \ll \rho_{\rm 2}$ and $c_{\rm s1}, c_{\rm s2} \ll 1$.

\section{The models}\label{themodels}
\setcounter{equation}{0}

SFS and FSFS show up within the framework of the Einstein-Friedmann cosmology governed by the
standard field equations
\bea \label{rho} \varrho(t) &=& \frac{3}{2}
\left(\frac{\dot{a}}{a} 
\right)^2~,\\
\label{p} p(t) &=& - \frac{1}{2} \left(2 \frac{\ddot{a}}{a} + \frac{\dot{a}^2}{a^2}  \right)~,
\eea
%
%
%
%
%
where the energy-momentum conservation law,
\be
\label{conser}
\dot{\varrho}(t) = - 3 \frac{\dot{a}}{a}
\left(\varrho(t) + p(t) \right),~
\ee
is trivially fulfilled due to the Bianchi identity. Here the dot means the derivative with respect to the time $t$. For $\mathcal{K}=0$ we can define an effective barotropic index as \cite{DDGH}
\be
\label{weff}
w_{eff}(t) = \frac{p}{\varrho } = \frac{1}{3} \left( 2 q(t) -1 \right),
\ee
where $q(t) = - \ddot{a} a/\dot{a}^2$ is the deceleration parameter. In both cases of SFS and FSFS singularities we take the scale factor in the form
\be
\label{sf2} a(t) = a_s \left[\delta + \left(1 - \delta \right) \left( \frac{t}{t_s} \right)^m - \delta \left( 1 - \frac{t}{t_s} \right)^n \right],
\ee
with the appropriate choice of the constants $\delta, t_s, a_s, m,n$ \cite{barrow04,DHD}. For both SFS and FSFS models described in terms of the scale factor (\ref{sf2}), the evolution begins with the standard big-bang
singularity at $t=0$, where $a=0$, and finishes at an exotic singularity for $t=t_s$, where $a=a_s\equiv a(t_s)$ is a constant. 

In order to have accelerated expansion in an SFS universe, $\delta$ has to be negative ($\delta<0$), and in an FSFS universe, $\delta$ has to be positive ($\delta>0$). For $1<n<2$ we have an SFS, while in order to have an FSFS, $n$ has to be in the range $0<n<1$. As it can be seen from (\ref{rho})-(\ref{sf2}), for SFS at $t=t_s$, $a \to a_s$, $\varrho \to \varrho_s=$ const., $p \to \infty$, while for an FSFS the energy density $\rho$ also diverges and we have: for $t\rightarrow t_s$, $a\rightarrow a_s$, $\rho\rightarrow\infty$, and $p \rightarrow \infty$, where $a_s,\ t_s$, are constants and $a_s\neq 0$. A special case of an SFS in which the anti-Chaplygin gas equation of state is allowed \cite{kamenshchik}
\be
p(t) = \frac{A}{\varrho(t)} \hspace{0.5cm} (A \geq 0)~~,
\label{eoschap}
\ee
is a big-brake singularity for which $\varrho \to 0$ and $p \to \infty$ at $t=t_s$ \cite{DDGH}. These special models have been checked against data in Refs. \cite{laszlo, sc2}.




Both SFS and FSFS scenarios consist of two components such as, a nonrelativistic matter, and an exotic fluid (which we will further call dark energy with the energy density $\rho_{de}$) which drives a singularity. We consider the case of the noninteracting components $\rho_m$ and $\rho_{de}$  which both obey independently their continuity equations of the type (\ref{conser}). The evolution of both ingredients is independent. Nonrelativistic matter scales as $a^{-3}$, i.e.
\be
\rho_m=\Omega_{m0}\rho_0\left(\frac{a_0}{a}\right)^3,
\ee
and the evolution of the exotic (dark energy) fluid $\rho_{de}$, can be determined by taking the difference between the total energy density $\rho$, which enters the  Friedmann equation (\ref{rho}), and the energy density of nonrelativistic matter, i.e.
\be
\rho_{de}=\rho-\rho_m~~.
\ee
In fact, $\rho_{de}$ component of the content of the Universe is responsible for an exotic singularity at $t\rightarrow t_s$. The dimensionless energy densities are defined in a standard way as
\be
\Omega_m=\frac{\rho_m}{\rho}, \hspace{0.3cm} \Omega_{de}=\frac{\rho_{de}}{\rho}.
\ee
For a dimensionless exotic dark energy density we have the following expression
\be
\label{Omde}
\Omega_{de}=1-\Omega_{m0}\frac{H_0^2}{H^2(t)}\left(\frac{a_0}{a(t)}\right)^3=1-\Omega_{m}.
\ee
We can then define the barotropic index of the equation of state for the dark energy as
\be
\label{wde}
w_{de}=p_{de}/ \rho_{de},
\ee
and the effective barotropic index of the total equation of state is given by Eq. (\ref{weff}).

\section{Perturbations}\label{perturbations evo}
\begin{figure}
 \begin{tabular}{cc}
  \resizebox{73mm}{!}{\includegraphics{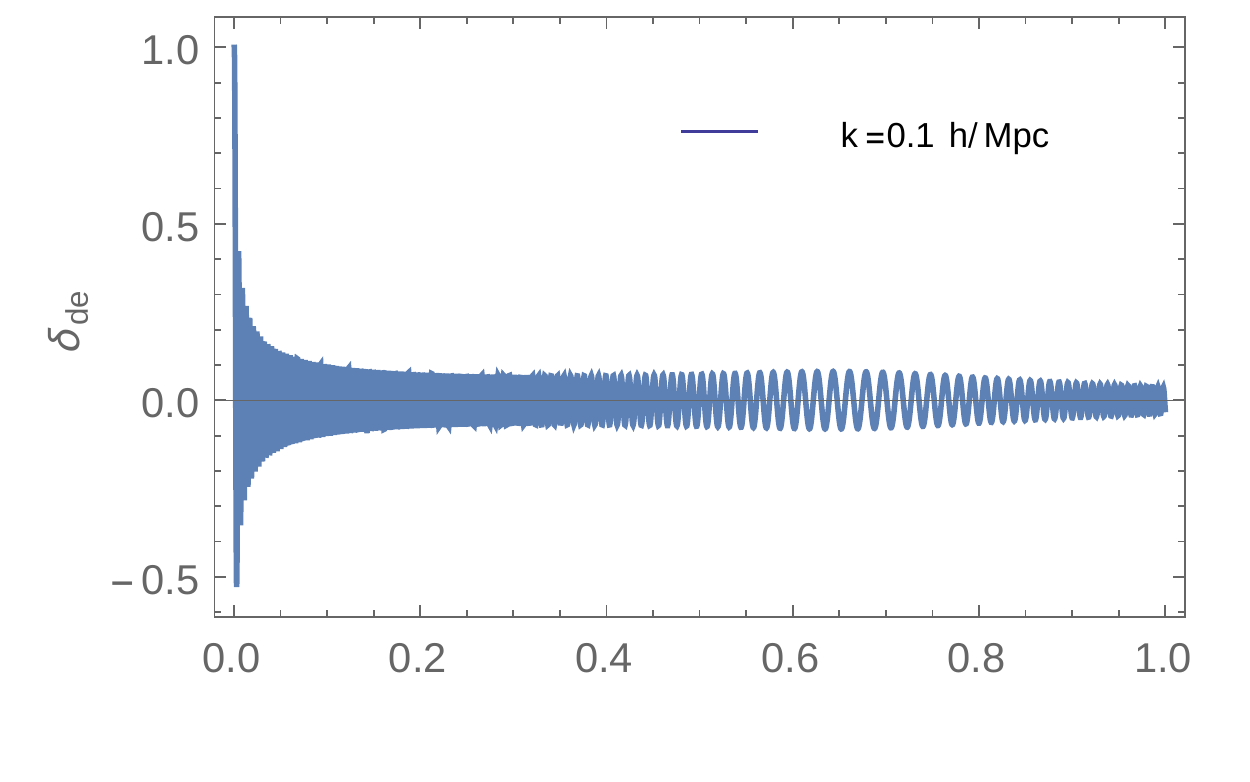}}&
  \resizebox{73mm}{!}{\includegraphics{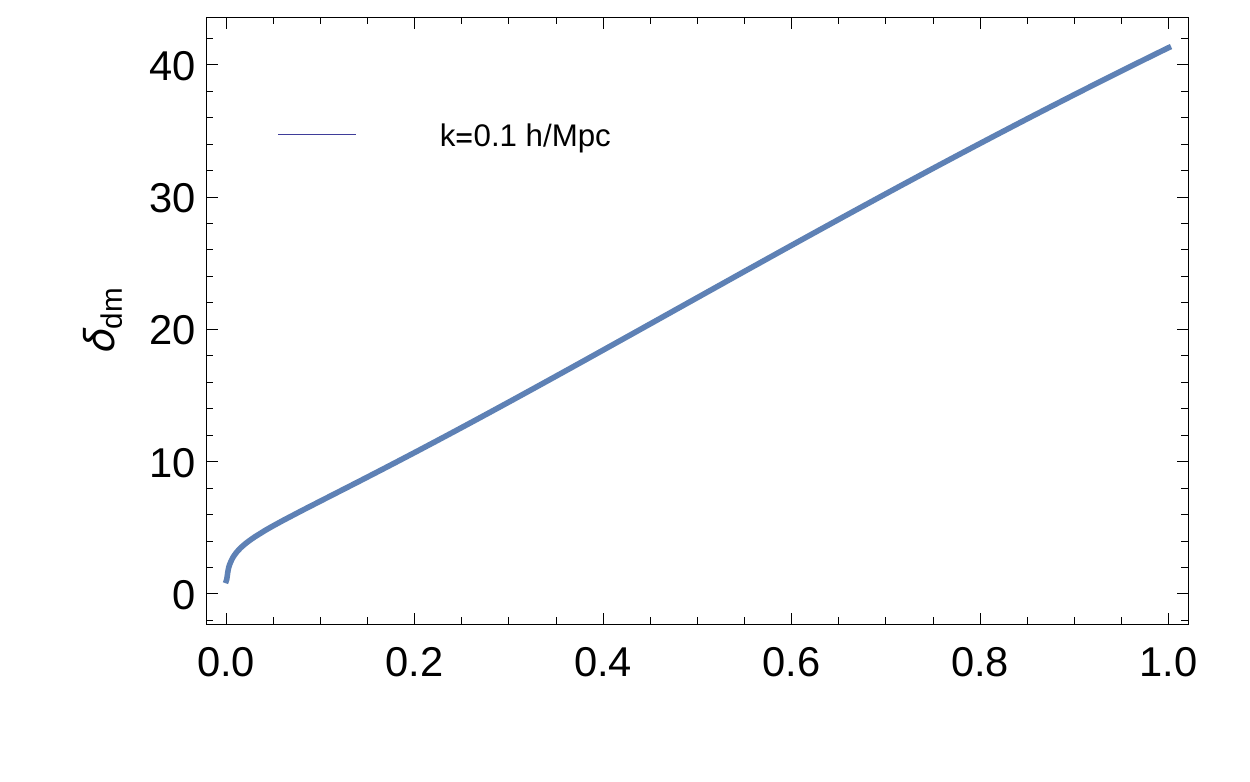}}\\
  \resizebox{73mm}{!}{\includegraphics{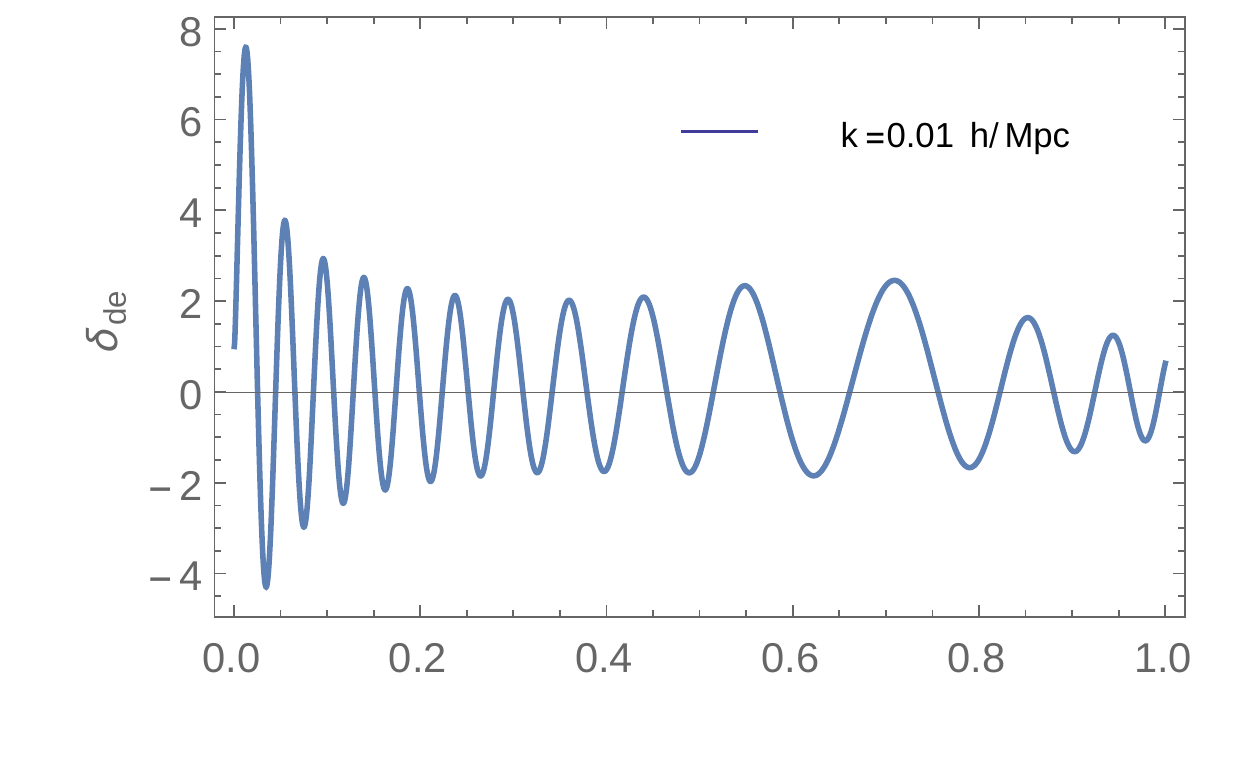}}&
  \resizebox{73mm}{!}{\includegraphics{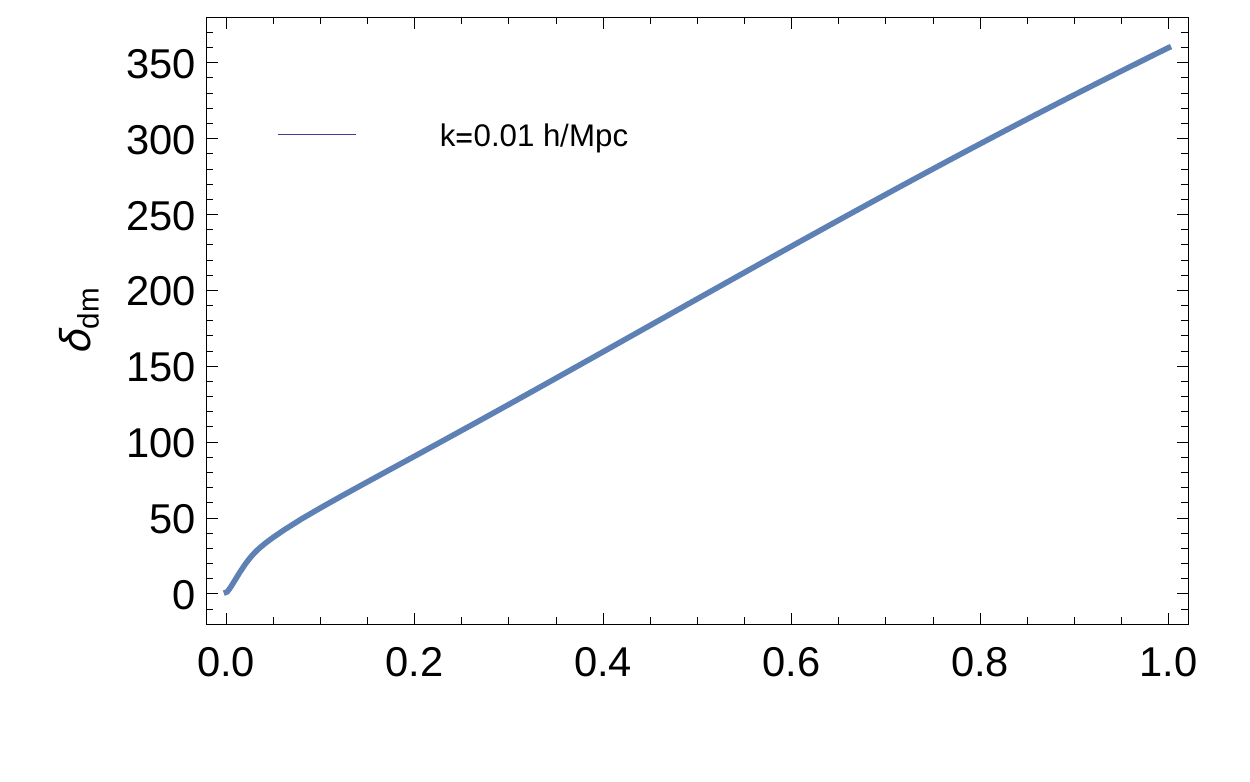}}\\
  \resizebox{73mm}{!}{\includegraphics{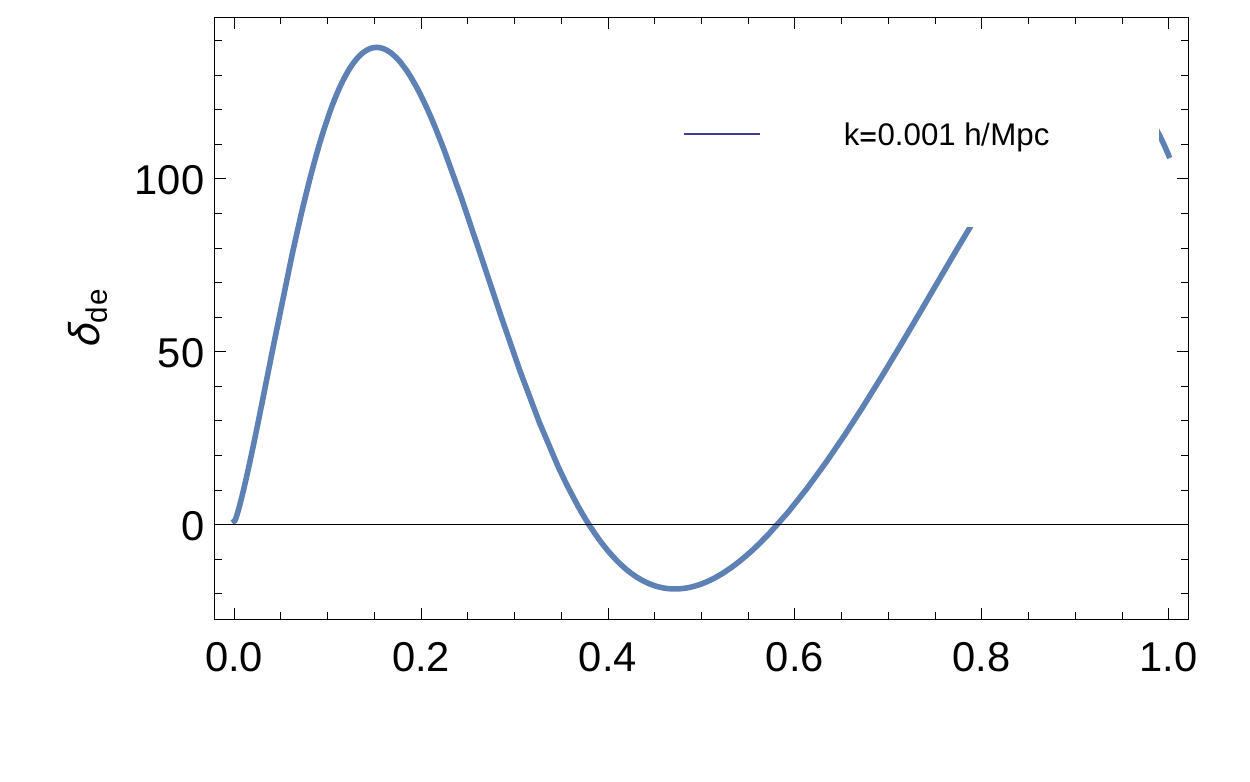}}&
  \resizebox{73mm}{!}{\includegraphics{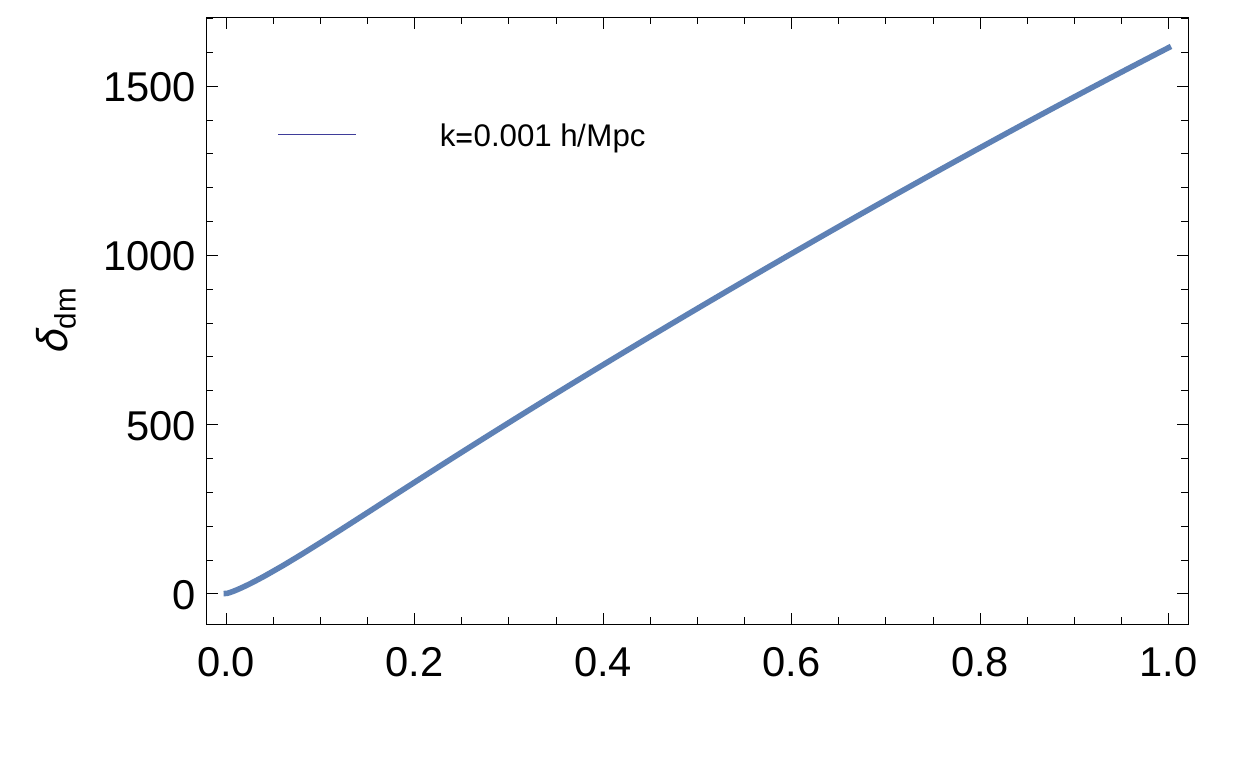}}\\
  \resizebox{73mm}{!}{\includegraphics{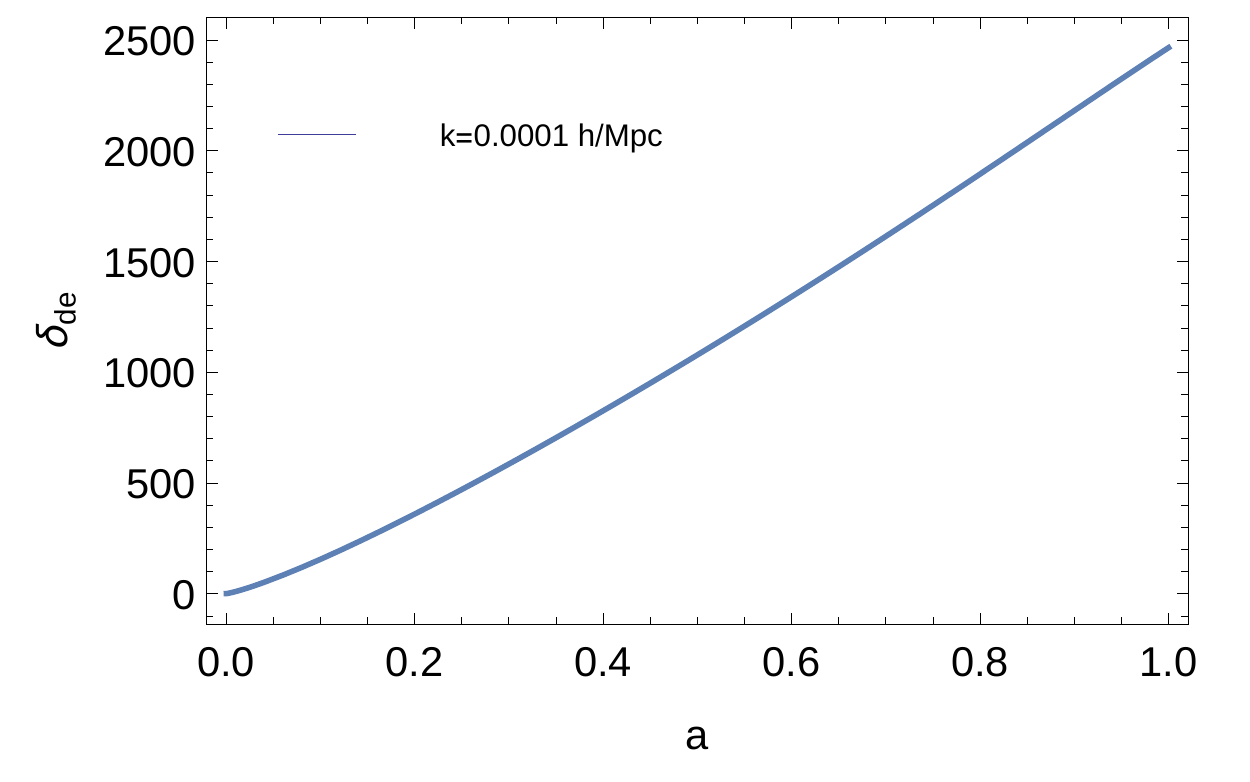}}&
  \resizebox{73mm}{!}{\includegraphics{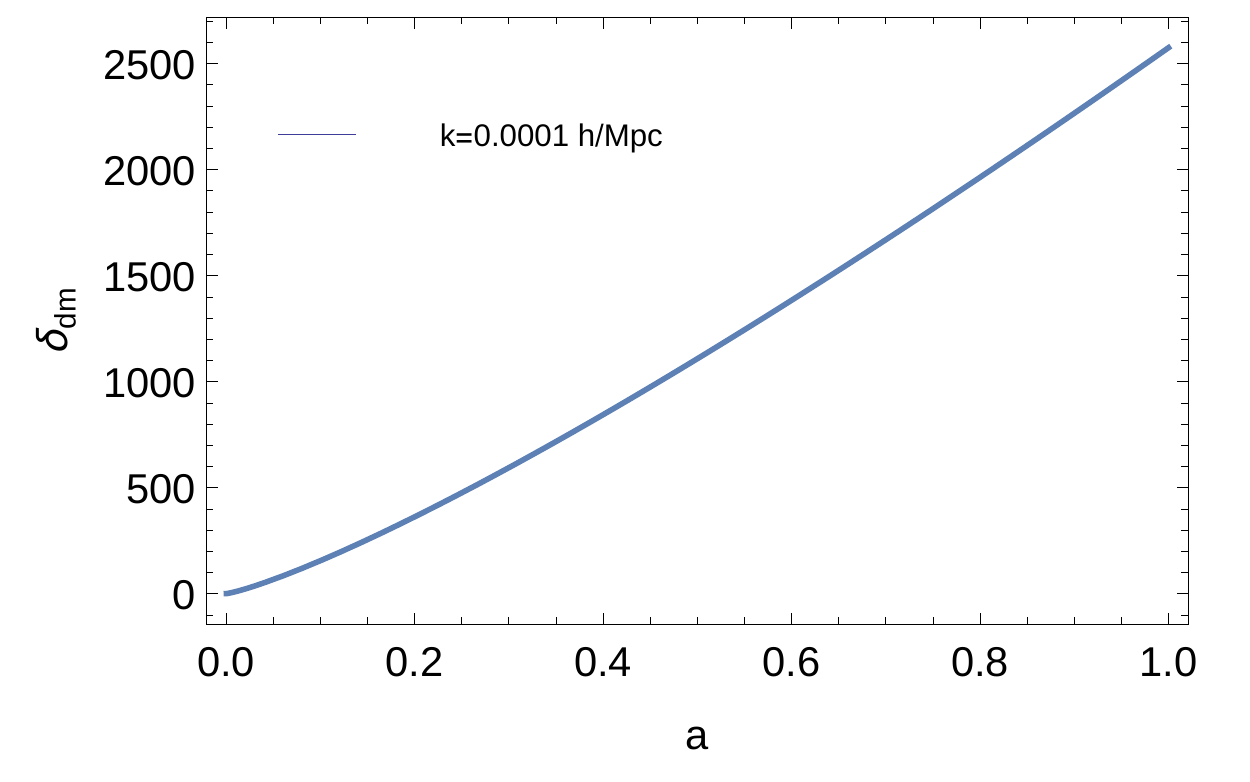}}\\
\end{tabular}

\caption{In the left column of the figure there are plots of dark energy density contrast evolution as a function of the scale factor. The scale factor takes the values within the range which corresponds to $z=1100$ and $z=0$ and is rescaled in a way that $a(t_0)=1$. The plots are for different values of the wavenumber given on each of the plots. The same plots are on the right column, but for matter density contrast. All plots are for the FSFS model with the values of the parameters $m=0.49$, $n=0.32$, $\delta=0.67$, $y0=0.55$, $\omega_m=0.315$.}\label{figuraa}
\end{figure}

\begin{figure}

 \begin{tabular}{cc}

  \resizebox{73mm}{!}{\includegraphics{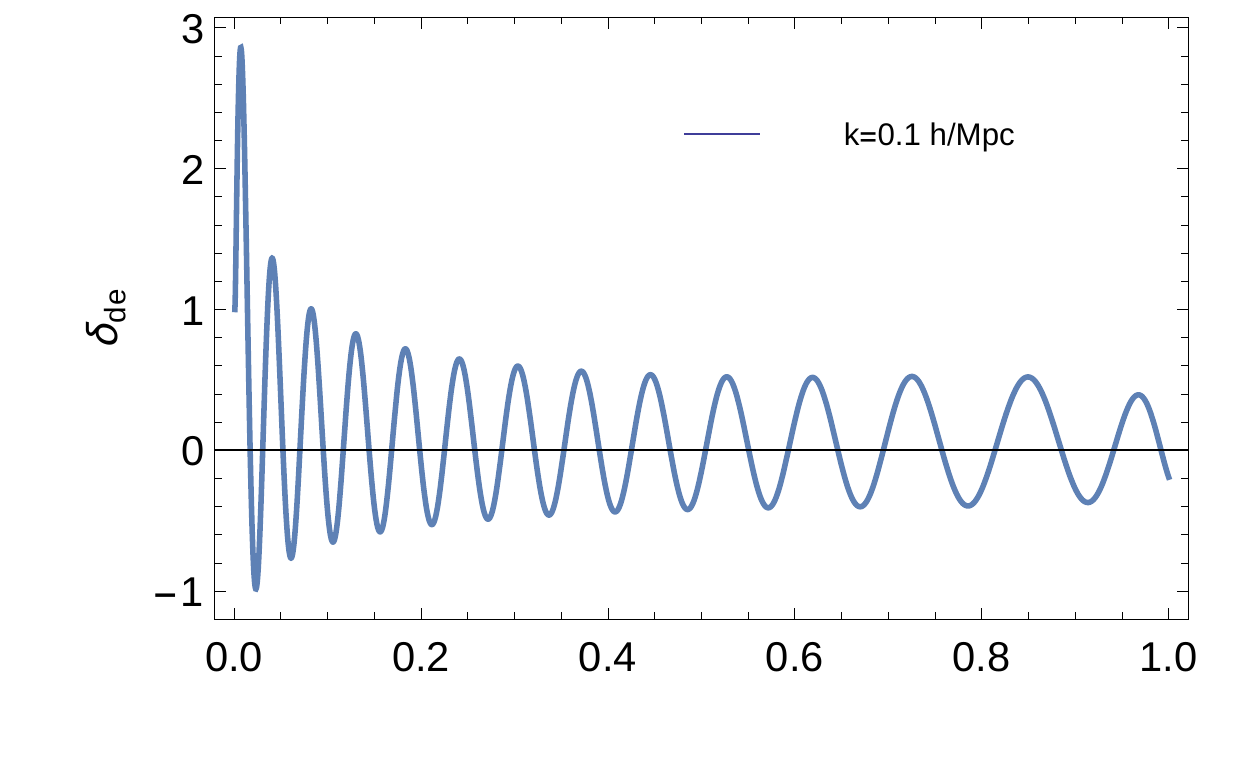}}&
  \resizebox{73mm}{!}{\includegraphics{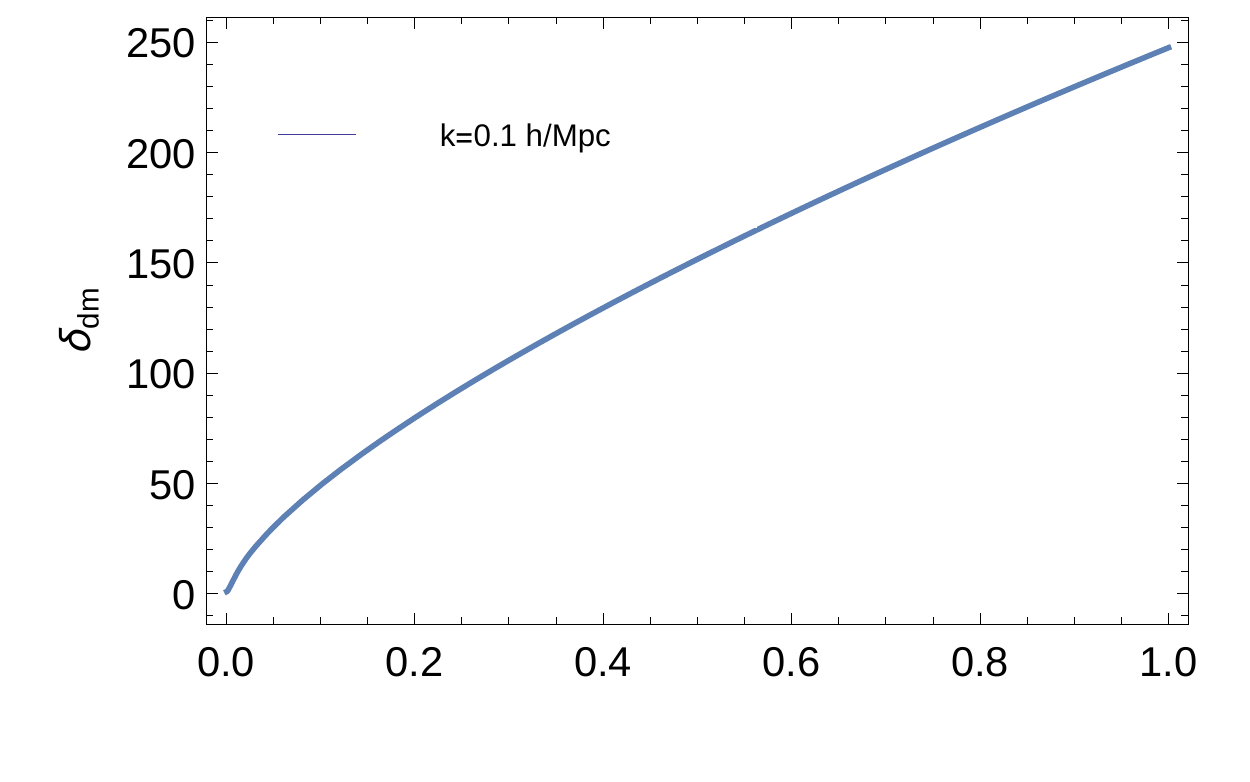}}\\
  \resizebox{73mm}{!}{\includegraphics{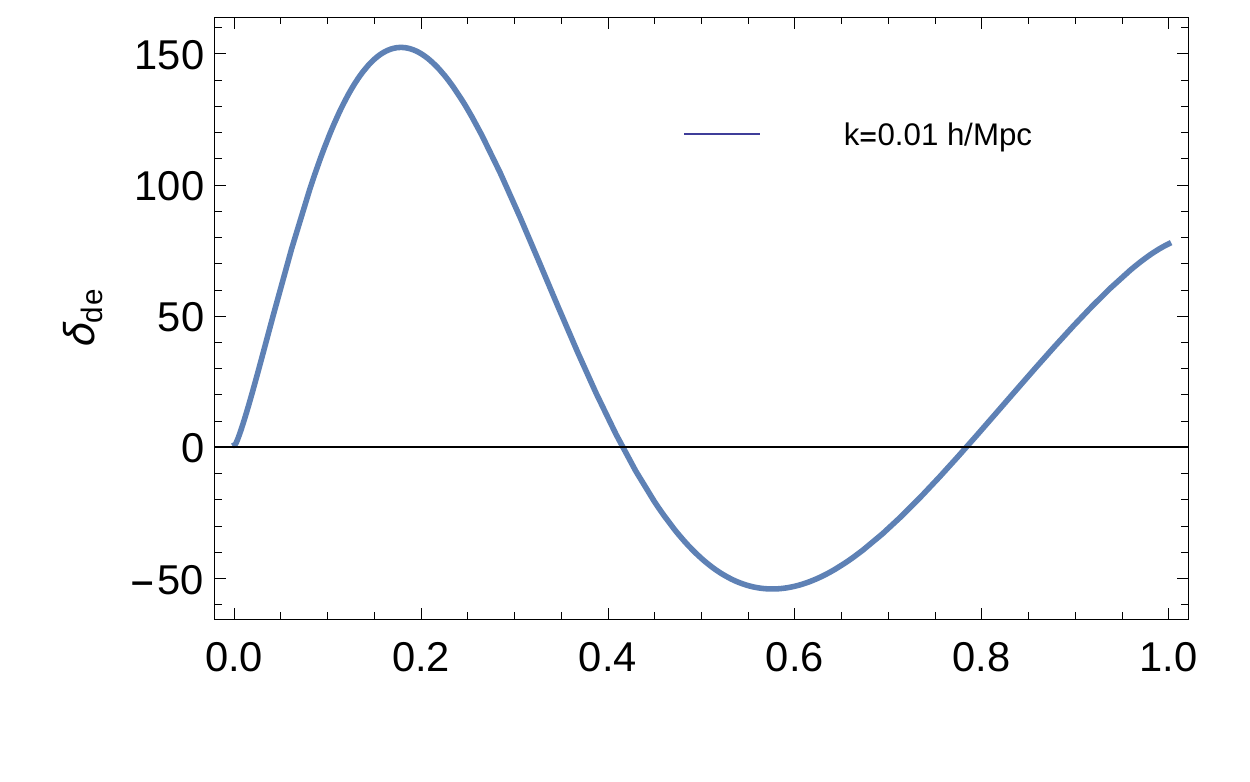}}&
  \resizebox{73mm}{!}{\includegraphics{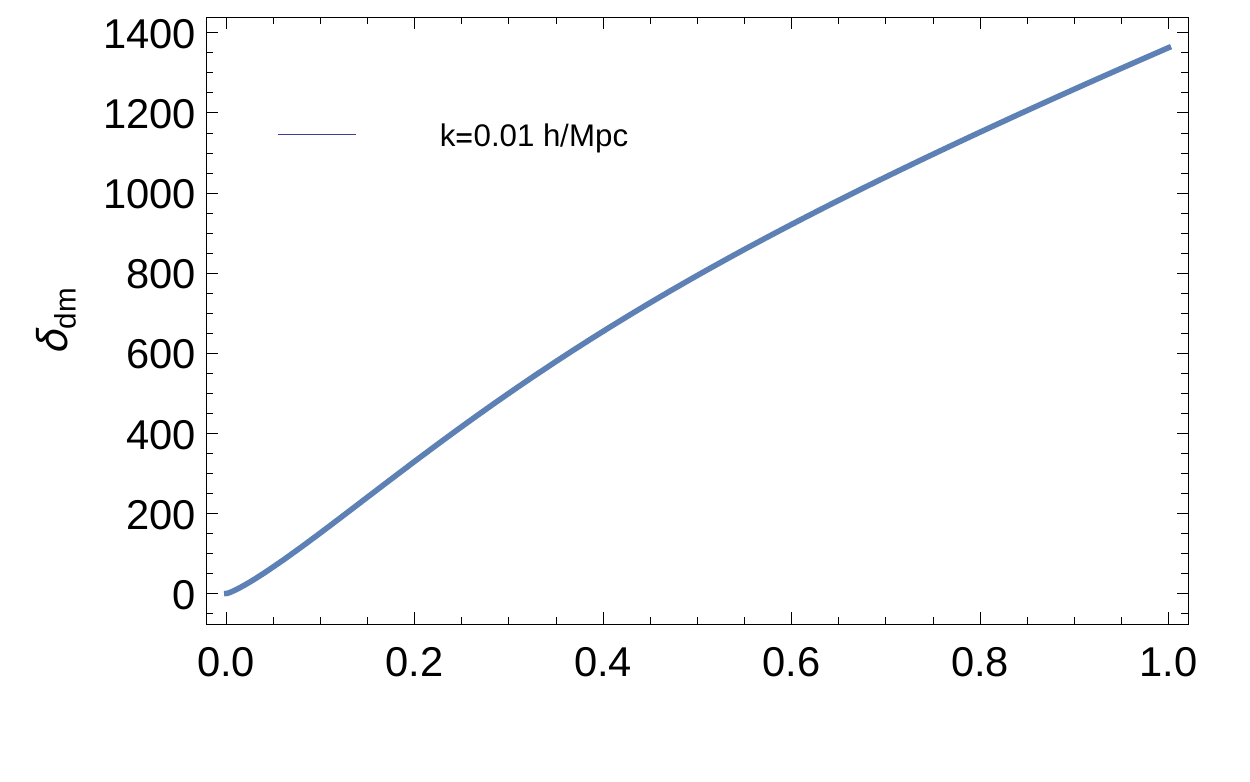}}\\
  \resizebox{73mm}{!}{\includegraphics{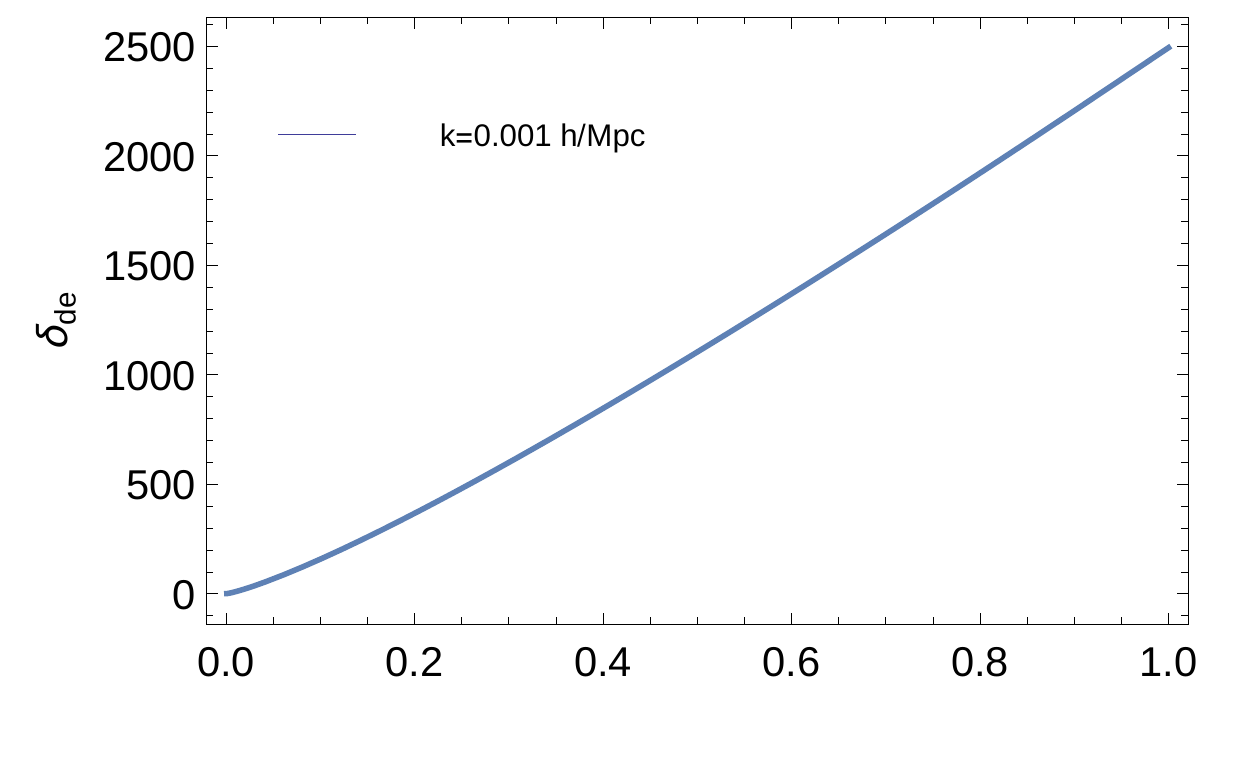}}&
  \resizebox{73mm}{!}{\includegraphics{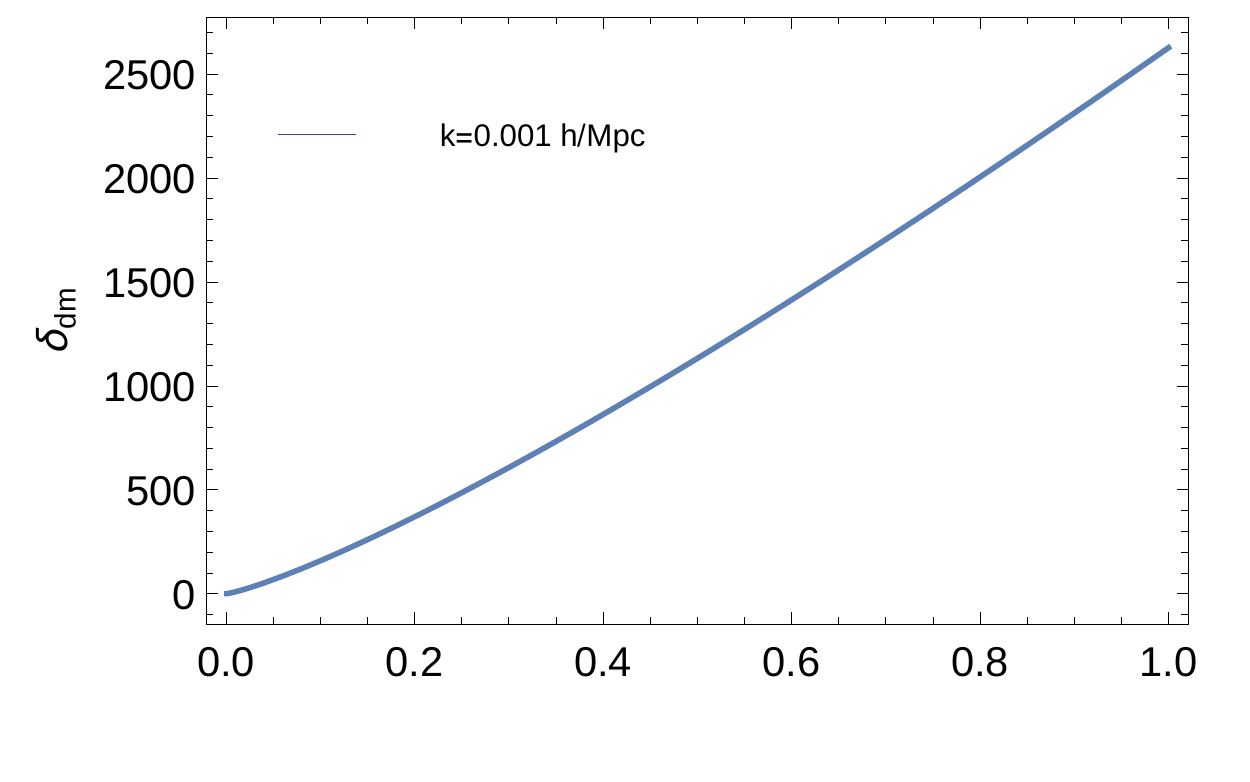}}\\
  \resizebox{73mm}{!}{\includegraphics{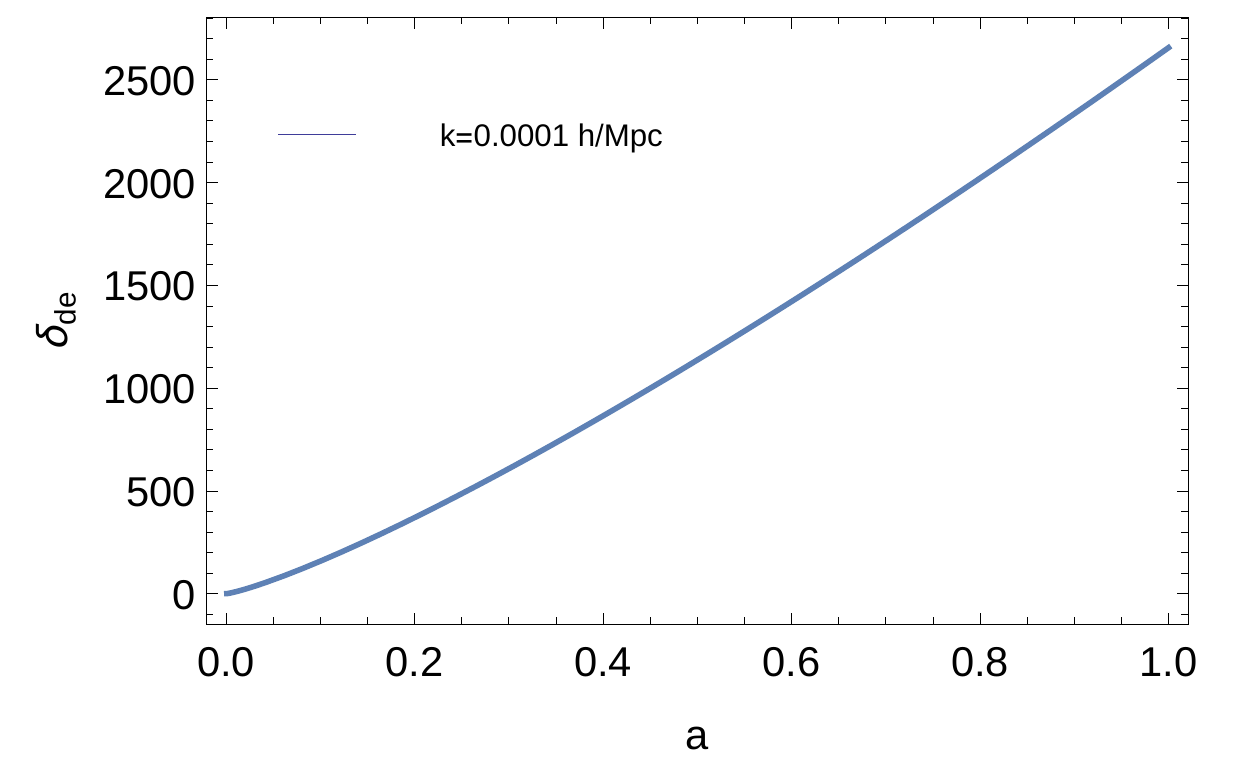}}&
  \resizebox{73mm}{!}{\includegraphics{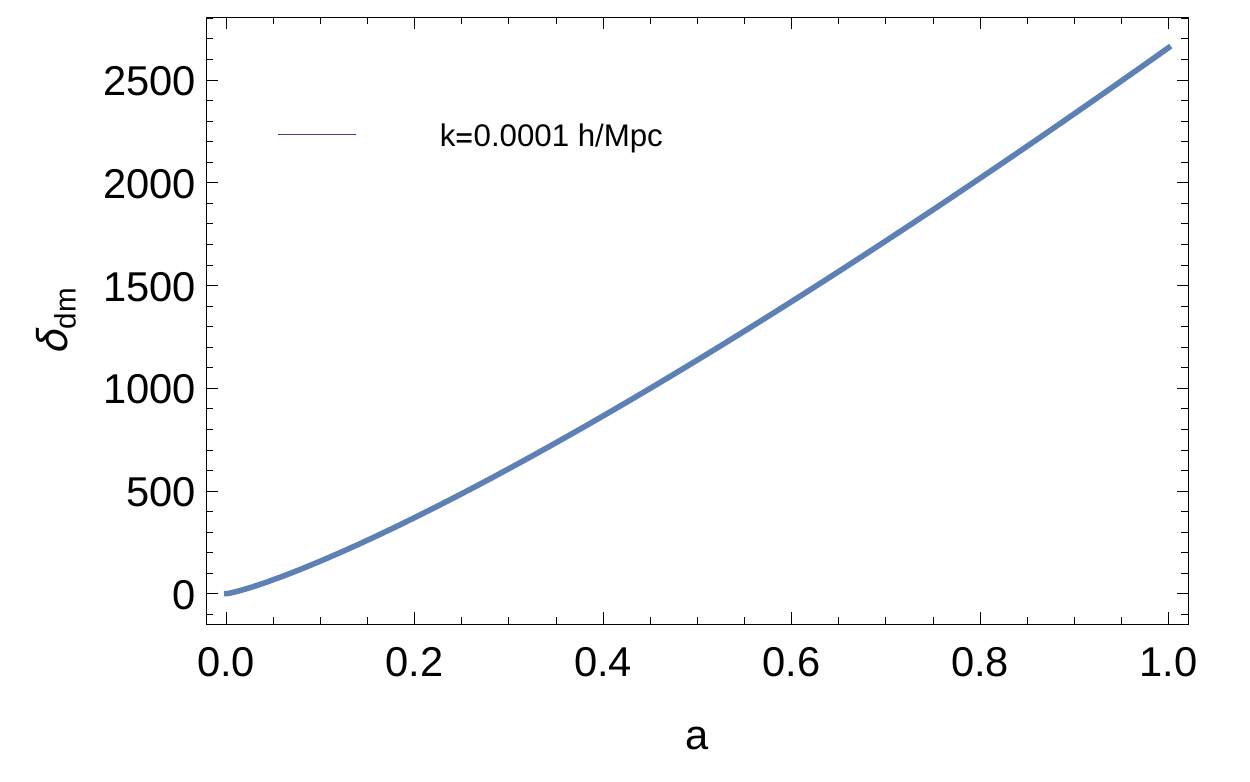}}\\

\end{tabular}
\caption{In the left column dark energy density contrast evolution as a function of the scale factor is plotted. Scale factor takes the values within the range which corresponds to $z=1100$ and $z=0$ and is rescaled in a way such that $a(t_0)=1$. The plots are for different values of the wavenumber given in each of the plots. The same plots are in the rhs column, but for the matter density contrast. All plots are for the SFS model with the following values of the parameters: $m=2/3$, $n=1.89$, $\delta=-0.15$, $y_0=0.995$, $\omega_m=0.315$.}\label{figurab}
\end{figure}

\begin{figure}

 \begin{tabular}{cc}
  \resizebox{73mm}{!}{\includegraphics{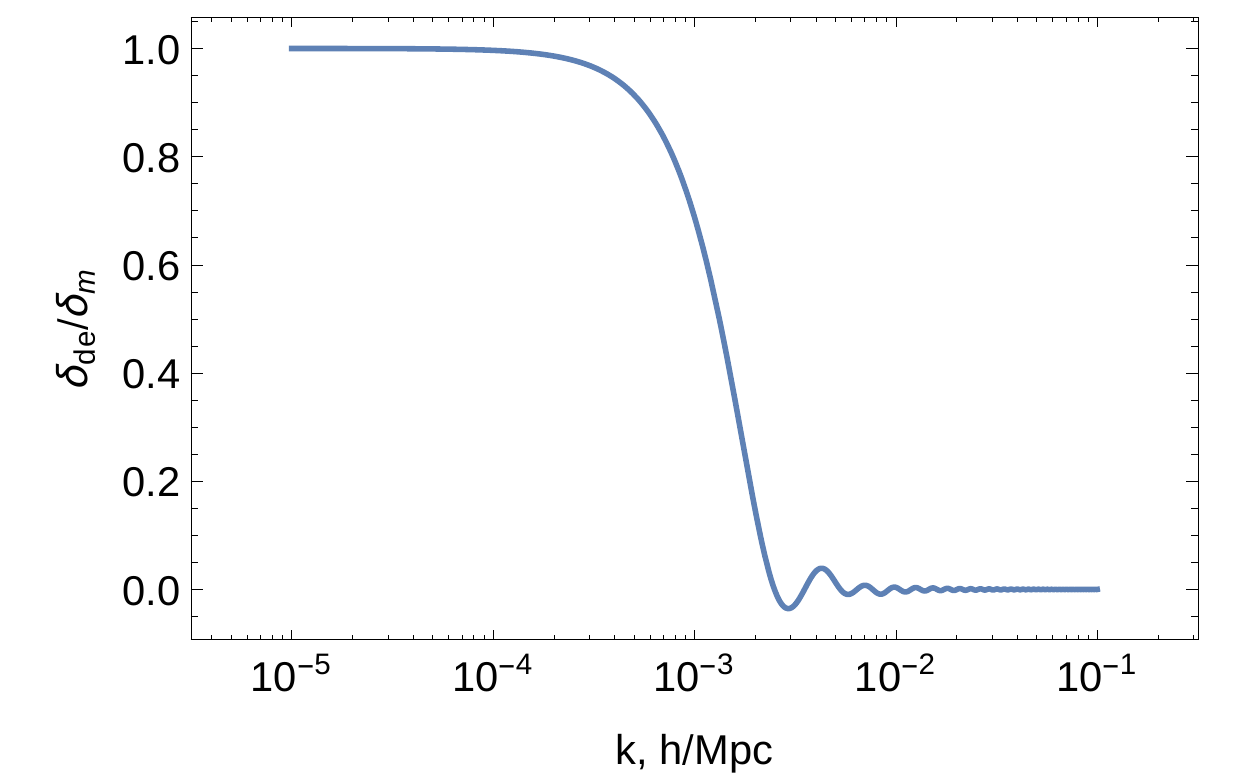}}&
  \resizebox{73mm}{!}{\includegraphics{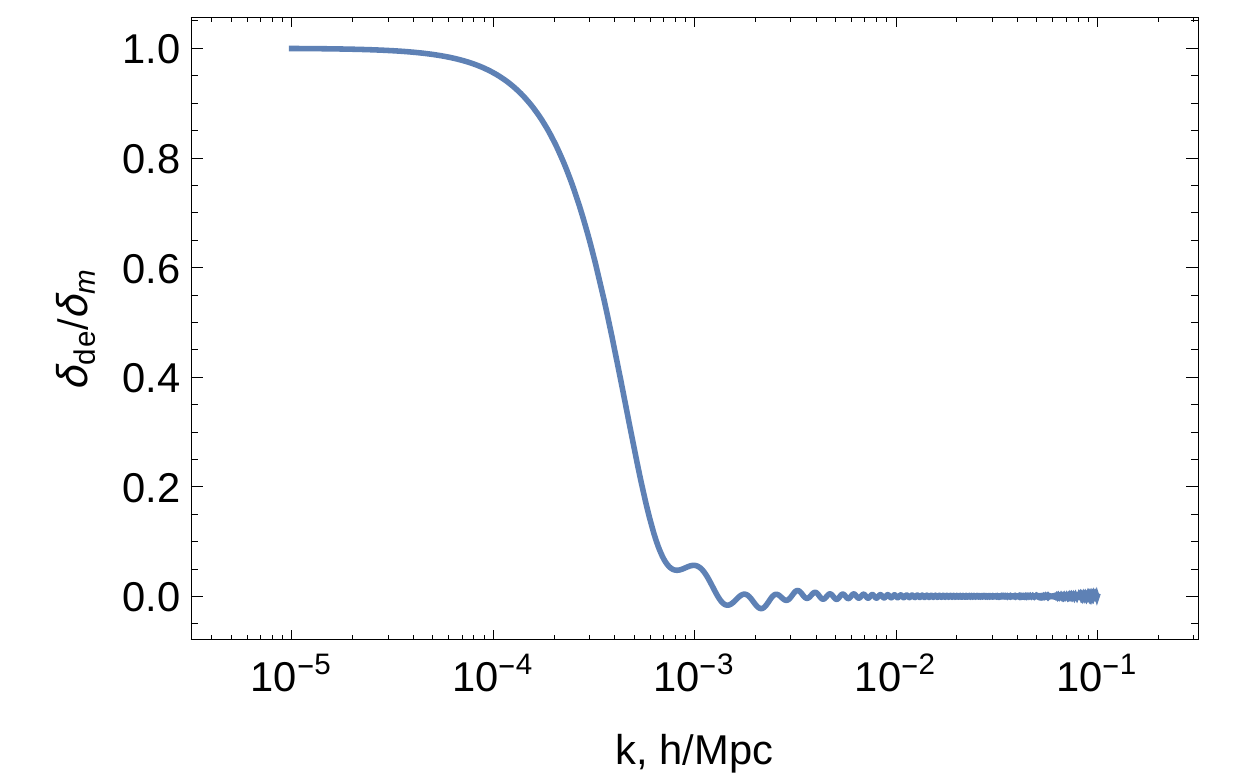}}\\
\end{tabular}
\caption{Plots of the density contrast ratio for dark energy and dark matter perturbations as a function of the wavenumber, at $z=0$; \underline{left}: the SFS model and \underline{right}: FSFS model with the values of the parameters mentioned earlier in Fig. \ref{figuraa} and Fig. \ref{figurab};}\label{dmoverde}
\end{figure}

The set of equations (\ref{simp}) for SFS and FSFS scenarios considered in this work take the following form:
\be\label{tosolve}
\begin{array}{c}
  \ddot{\delta}_{\rm m}+\left(\frac{\dot{a}}{a}-\frac{\ddot{a}}{a}\right)\dot{\delta}_{\rm m \ }+\frac{k^2}{a^2}c_{s\rm m }^2\delta_{\rm m}=\rho_{\rm m}\delta_{\rm m}+\rho_{\rm de}\delta_{\rm de },\\
  \ddot{\delta}_{\rm de}+\left(\frac{\dot{a}}{a}-\frac{\ddot{a}}{a}\right)\dot{\delta}_{\rm de}+\frac{k^2}{a^2}c_{s\rm de}^2\delta_{\rm de}=\rho_{\rm m}\delta_{\rm m}+\rho_{\rm de}\delta_{\rm de},
\end{array}
\ee
The equations (\ref{tosolve}) are integrated with initial conditions $\{\delta_{dm}(z_{rec}),$ $\dot{\delta}_{dm}(z_{rec}),$ $ \delta_{de}(z_{rec}),$ $\dot{\delta}_{de}(z_{rec})\}=$ $\{ 1,1,1,1\}$. We start the integration from the point around the recombination era, that is at $z_{rec}=1100$. The wavenumber is always given in units $[k]=hMpc^{-1}$ and is related to the wavelength of the perturbation by $\lambda=2\pi/k,$ with the appropriate units $[\lambda]=h^{-1}Mpc$.\\
In Fig. \ref{figuraa} we plot dark energy (lhs) and matter perturbations (rhs) evolution for a given wavenumber for FSFS cosmology with the values of the parameters given in the caption of the figure. The evolution of perturbations is plotted in the order, that the shortest wavelength of perturbations is at the top and largest scales are at the bottom. The amplitudes of the perturbations are plotted as a function of the scale factor.  In Fig. \ref{figurab} the same for SFS cosmology is plotted. In the case of evolving dark energy barotropic fluid, the signatures of dark energy perturbations should be expected at large scales (see also Fig. \ref{dmoverde}). On small scales, due to the fact that the modes of perturbations, in  the dark energy, enter the sound horizon early, they start to oscillate and eventually are suppressed. 

\begin{figure}
 \begin{tabular}{cc}

  \resizebox{73mm}{!}{\includegraphics{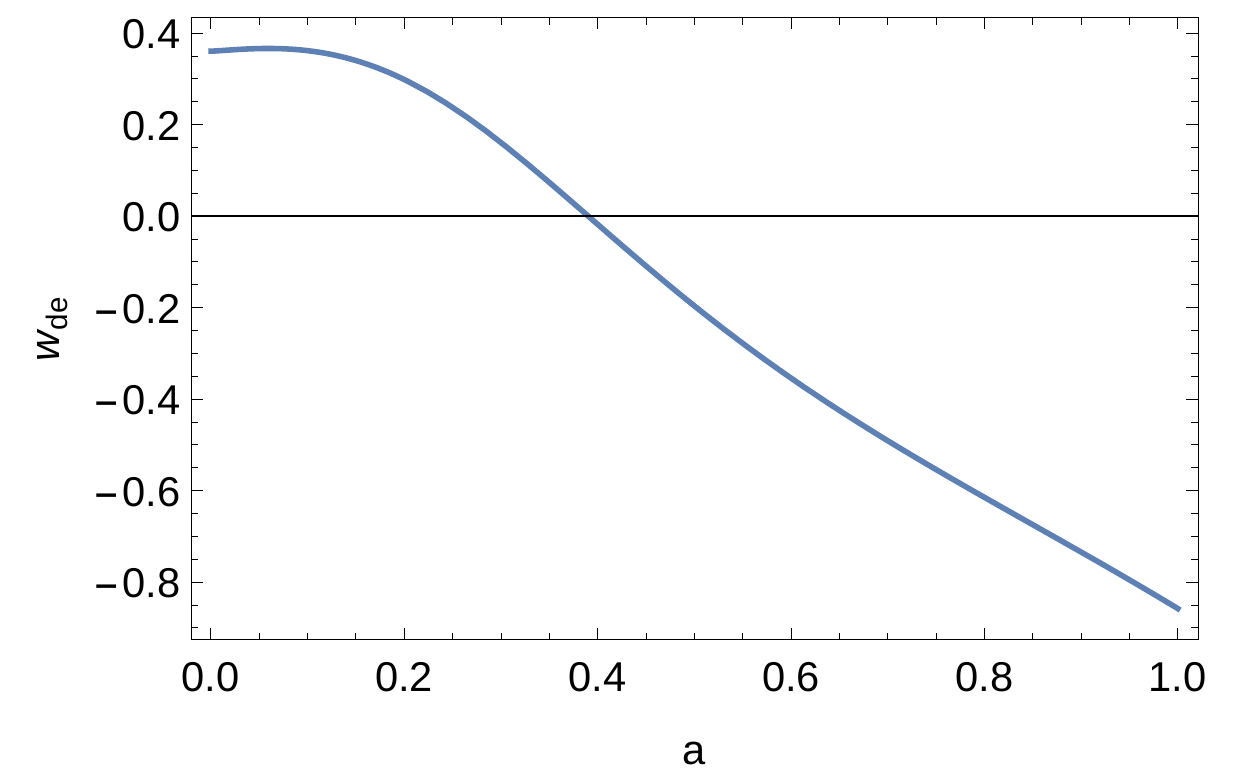}}&
  \resizebox{73mm}{!}{\includegraphics{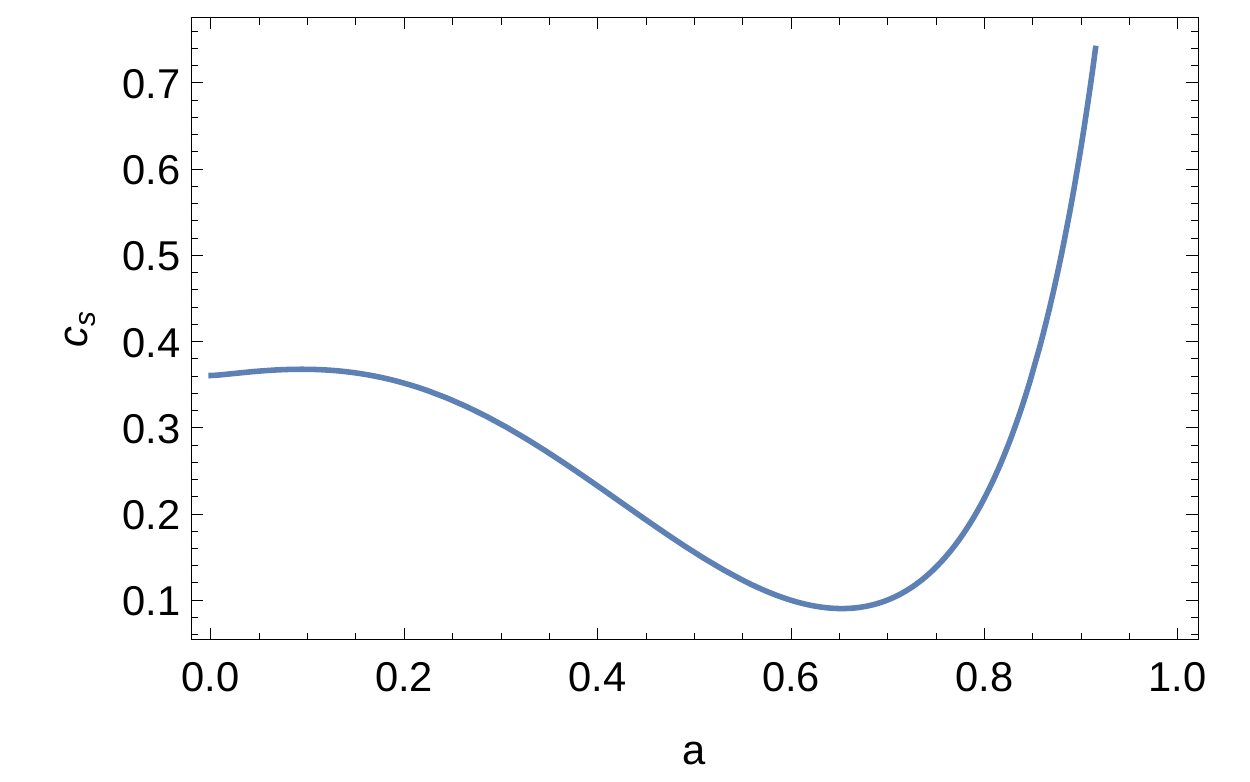}}\\
\end{tabular}

\caption{The plots for FSFS model with parameters given in Fig. \ref{figuraa}; \underline{left}: evolution of the dark energy equation of state parameter as a function of the scale factor; \underline{right}: sound speed evolution as a function of the scale factor; in both cases the scale factor ranges from the value corresponding to $z=1100$ to the value corresponding to $z=0$;}\label{wcsfsf}
\end{figure}

\begin{figure}

 \begin{tabular}{cc}
  \resizebox{73mm}{!}{\includegraphics{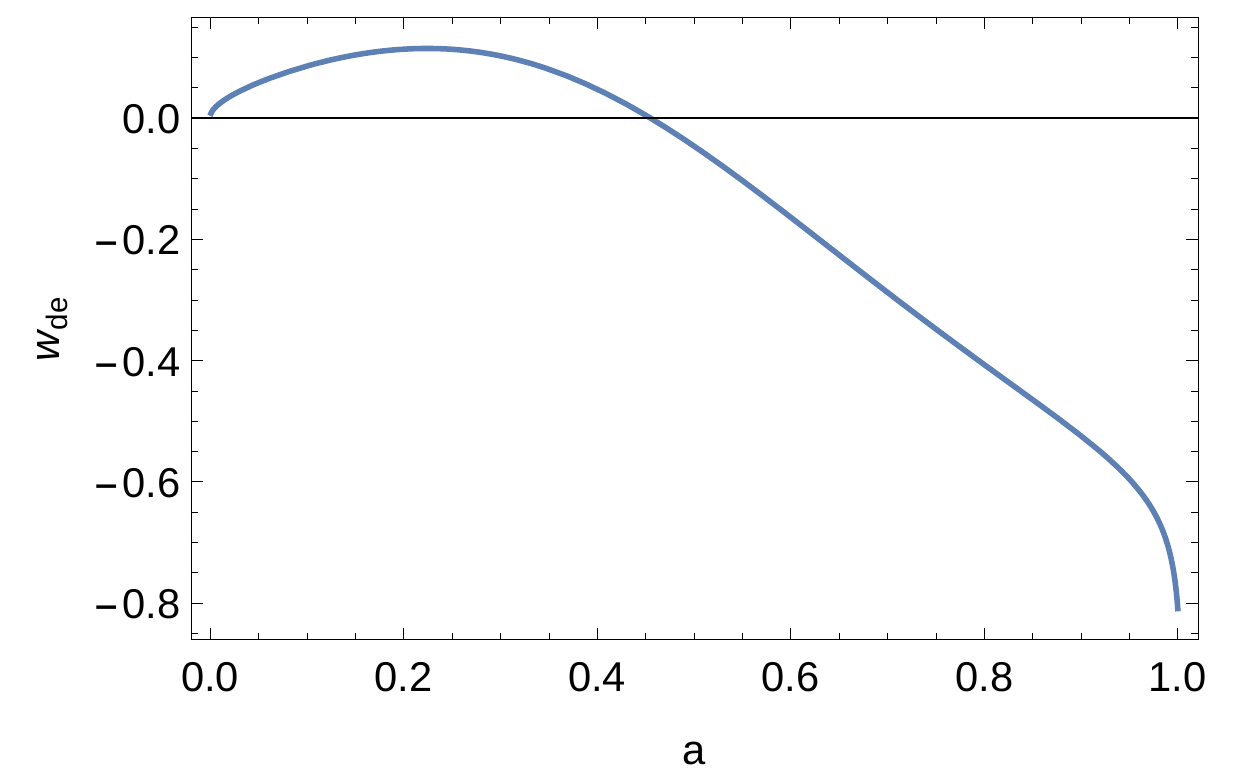}}&
  \resizebox{73mm}{!}{\includegraphics{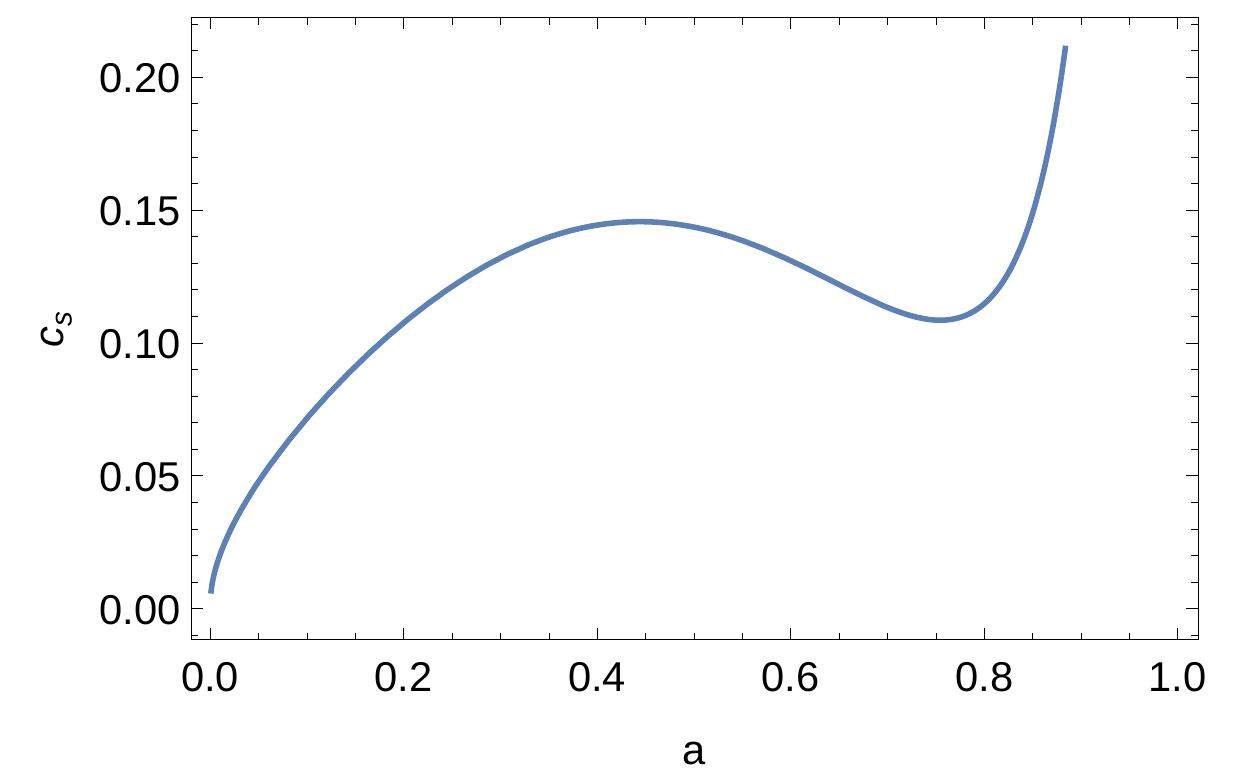}}\\
\end{tabular}
\caption{Plots for the SFS model with parameters given in Fig. \ref{figurab};  \underline{left}: the evolution of the dark energy equation of state parameter as a function of the scale factor; \underline{right}: sound speed evolution as a function of the scale factor;  in both cases the scale factor ranges from the value corresponding to $z=1100$ to the value corresponding to $z=0$; }\label{wcssfs}
\end{figure}

In the right panels of Figs. \ref{wcsfsf} and \ref{wcssfs} the sound speed as a function of the scale factor is plotted for the FSFS and SFS cosmologies, respectively. As suggested in \cite{1202.3280v2}, when we restrict our investigations to some certain cases, dark matter perturbations effectively decouple from perturbations at dark energy. Conditions are the following: the sound speed for the dark energy has a positive value of order unity, the barotropic index for the dark energy is a reasonably slowly varying function of the cosmic time and the perturbation mode is well within the Hubble horizon.  Fast oscillating dark energy perturbations do not influence growth of matter perturbations within the wide range of wavelengths. The evolution  of the matter density contrast $\delta_m$ can be described to a good approximation with the equation for scale independent growth rate,
\be
f^\prime+f^2+f\left(\frac{\dot{H}}{H^2}+2\right)-\frac{3}{2}\Omega_m=0,
\ee
where $f=d\ln \delta /d\ln a$ and $\prime=d/d\ln a$.

\begin{figure}

 \begin{tabular}{cc}
  \resizebox{73mm}{!}{\includegraphics{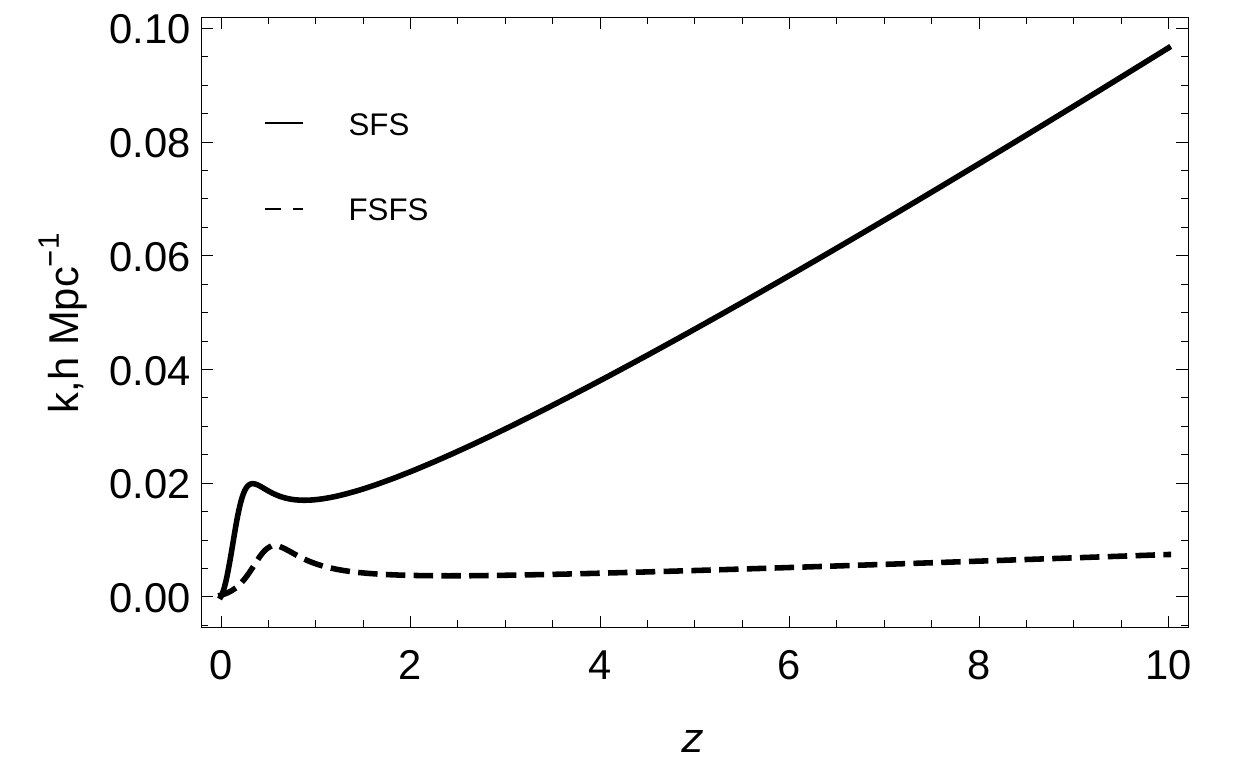}}&
  \resizebox{73mm}{!}{\includegraphics{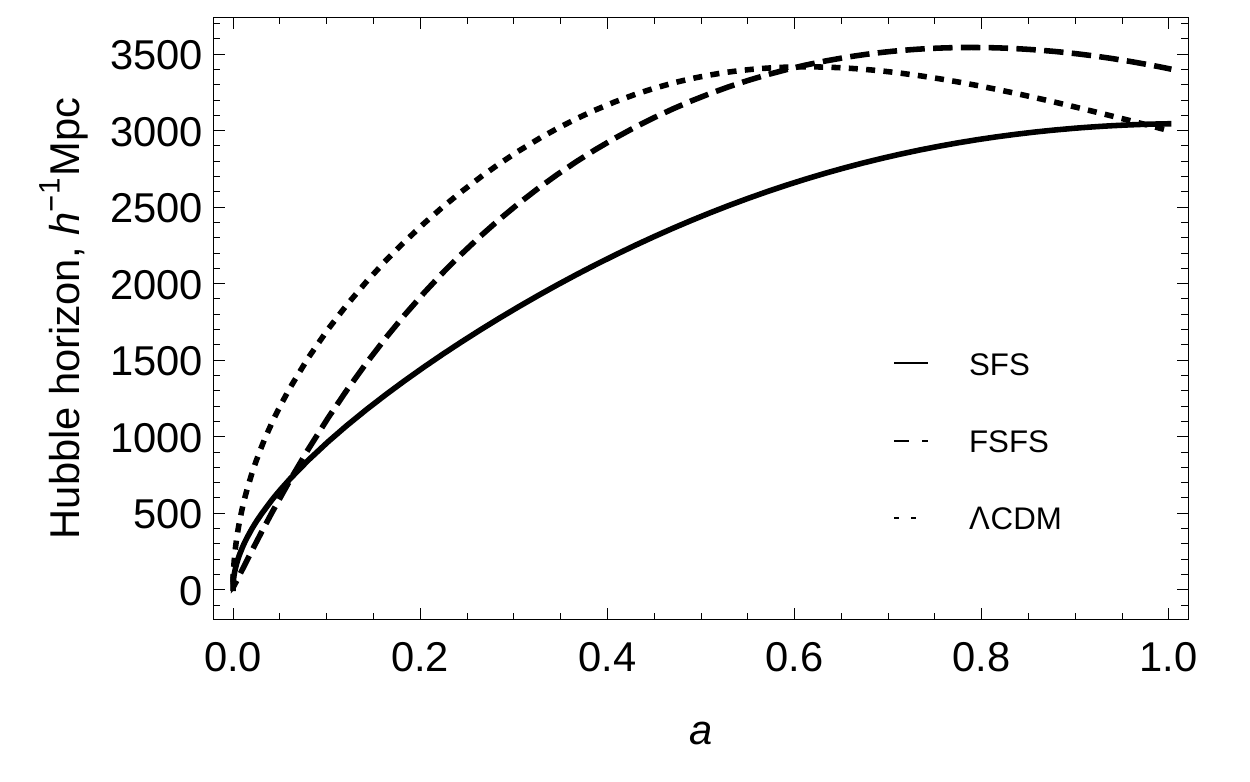}}\\
\end{tabular}
\caption{\underline{left}: the value of the wavenumber which corresponds to the perturbation modes which are within sound speed horizon as a function of redshift, for the SFS and FSFS models; \underline{right}: the Hubble horizon for SFS, FSFS, and $\Lambda$CDM models as a function of the scale factor; the scale factor ranges from the value corresponding to $z=1100$ to the value corresponding to $z=0$ and is rescaled such that $a_0=a(t_0)=1$;
}\label{horizons}
\end{figure}

At the left panel of Fig. \ref{horizons} the value of the wavenumber, which corresponds to the perturbation modes, which are within sound horizon at a given redshift, for FSFS and SFS cosmologies are plotted. 
For both models, modes of the dark energy perturbations start to grow after they enter the Hubble horizon. On the rhs of the Fig. \ref{horizons}, the Hubble horizon as a function of the scale factor, for each model, is plotted. Eventually, when perturbation modes enter the sound horizon, the wavenumber $k$ satisfies the relation, $k>aH/c_s$. The behaviour of the evolution of the growth, its frequency of the oscillations, and the damping is determined by the relation between the wavenumber, $k$ and the sound horizon $aH/c_s$. That can be seen from Eq. (\ref{simp}). The earlier the wavelength of the mode is comparable to the sound horizon the sooner the mode is damped and the perturbations start to oscillate and are then suppressed. The modes with larger wavenumbers enter the sound horizon earlier, at larger redshifts/smaller scale factor and their oscillation and damping starts earlier.

In Fig. \ref{dmoverde} the ratio of the dark energy to the matter perturbations $\delta_{\rm de}/ \delta_{\rm dm}$, as a function of the wavenumber for FSFS and SFS  is plotted.  For small length scale through intermediate length scales the perturbations in dark energy can be neglected. But on large scales or small wavenumber, $k<2\cdot 10^{-4}h/Mpc$ for FSFS and $k<7\cdot 10^{-4}$ for SFS, dark energy perturbations can acquire the values of the order of the dark matter perturbations.  

In the framework that we adopted in this paper in the case of barotropic index crossing the $w=-1$ divide, the perturbations would encounter a blowup. We consider the models with dark energy equation of state which is positive at the beginning and do not cross $w=-1$ divide. For both models the present value of barotropic index for dark energy is $w_{de}\simeq-0.8$, which is in agreement with the Planck result \cite{planck13}.

\begin{figure}
\begin{center}
  \resizebox{73mm}{!}{\includegraphics{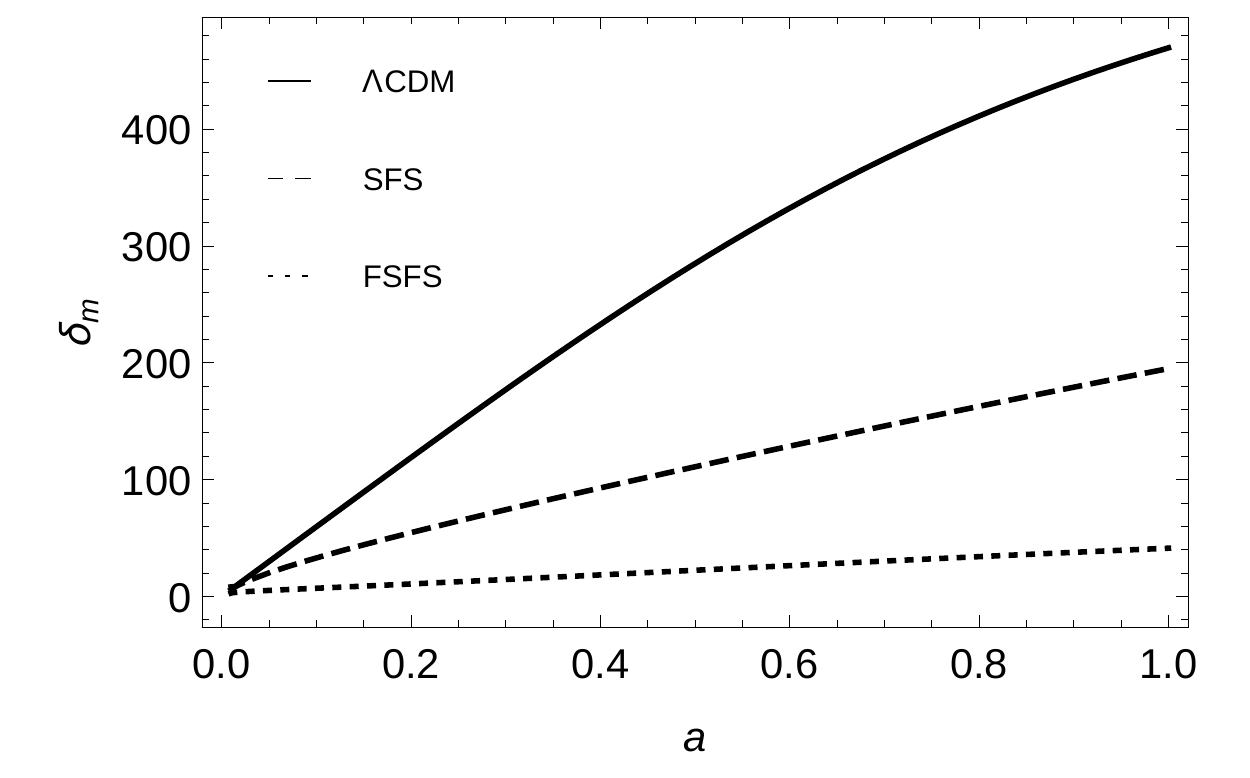}}
\caption{Plot of the dark matter contrast evolution for FSFS, SFS and $\Lambda$CDM model for the mode with the wavenumber, $k=0.1 h/Mpc$. Values of the parameters for the models with future singularity are taken as previously mentioned; for $\Lambda$CDM consistently $H_0=70km/s/Mpc$ and $\Omega_m=0.315$; the scale factor ranges from the value corresponding to $z=1100$ to the value corresponding to $z=0$ and is rescaled such that $a_0=a(t_0)=1$. }\label{lcdmt23del}

\end{center}

\end{figure}

In Fig. \ref{lcdmt23del} the evolution of the dark matter density contrast for $\Lambda$CDM and singular models for the wavenumber $k=0.1hMpc^{-1}$ is plotted. Although singular models considered here satisfy geometrical probes, there is the significant difference in the growth of the matter perturbations between the models and $\Lambda$CDM, as well as, between the models themselves.   



\section{Conclusion}\label{konk}
\setcounter{equation}{0}
We have used gauge invariant formalism for the evolution of perturbations within the models of the universe, which exhibit singular behaviour of pressure only or pressure and the energy density in a finite time. We explored particular models of finite scale factor singularity (FSFS) and sudden future singularity (SFS). We analyzed the behaviour of perturbations of the dark matter and dark energy for several wavelengths as functions of the scale factor. We plotted the Hubble horizon, barotropic index for dark energy equation of state, and the speed of sound as a functions of the scale factor. We also investigated ratio of dark energy and dark matter perturbations as a function of the wavenumber.  

At the moment with constraints coming from the geometric probes of the universe it is not possible to rule out finite future singularities or discriminate between those models. Possibly future data like the redshift drift \cite{rd} planned to be measured by ELT-HIRES \cite{eelt} will change the situation. Both models seem to be good candidates for evolving dark energy component with the ability to mimic $\Lambda$CDM model up to now. 

On the other hand, forthcoming future observations, such as  planned galaxy week lensing experiments like the Dark Energy Survey (DES) \cite{des} and Euclid \cite{euclid}, will potentially be able to discriminate between those models and $\Lambda CDM$. The characteristic features of the evolution of the matter and dark energy perturbations strongly depend on parameters. There is a chance to discriminate in the future between different types of singular cosmologies or even reject them as physical models. At least there is a possibility to put more stringent constraints onto models parameters. Also, the experiments to further explore cosmic background radiation in microwave to far-infrared bands in polarization and amplitude as PRISM \cite{prism} are planed. Very high precision measurements of the polarization of the microwave sky are planned to be done with CoRE \cite{core}. In view of this forthcoming data it is useful to explore perturbations in the dark sector to enhance its discriminating power.

The detailed analysis shows the differences in the evolution of the sound horizon during the course of the evolution of universe. This quantity determines the behaviour of the perturbations. The sooner the dark energy perturbation modes enter the sound horizon, the earlier they stop growing and start to oscillate and are suppressed. 

There are differences in the ratios of dark energy to dark matter perturbations as a function of the wavelength. For example, for the models explored here the dark energy perturbations become comparable to dark matter perturbations for shorter wavelengths $\lambda\simeq9000Mpc/h$ for the SFS scenario. For the FSFS wavelengths at which modes of dark matter and dark energy contrasts start to be comparable are: $\lambda\simeq30000Mpc/h$. There is a difference of the order of a magnitude between the models. As a result of differences in the Hubble and sound horizon evolution, there are differences in the dark energy perturbations evolution. If any, signal from perturbations of the evolving dark energy, should be seen for large scales. The scale on which dark energy perturbations become comparable with dark matter perturbations depend strongly on the model and its parameters.

From the plot of the dark matter density contrast for the reference $\Lambda$CDM model and singular scenarios in Fig. \ref{lcdmt23del}, it can be seen that the growth of perturbations in singular models differs from $\Lambda$CDM scenario and that there are differences between singular models due to different values of the parameters in the scale factor. This is encouraging for further investigation of those models with dynamical probes.  

On the other hand, for the modes which are well within Hubble horizon the dark matter perturbations effectively decouple from perturbations in the dark energy. In Figs. \ref{figuraa} and \ref{figurab} we plotted fast oscillations of dark energy perturbations within a wide range of wavelengths. For that reason, in that limit, dark energy perturbations do not influence growth of matter perturbations and the evolution of the matter density contrast $\delta_m$ can be described to a good approximation with the equation for scale independent growth function. 

\newpage
\section{Acknowledgements}
\indent Author acknowledges the support of the state budget for education grant No 0218/IP3/2013/72 (years 2013-2015).
\appendix
\section{Appendix}\label{app}
The coefficients in the equations (\ref{gensys}) have the following form \cite{Starobinski}:
\bea\label{abcd}
A_{\rm 1} = 2\frac{\h}{a} + 3\frac{\h}{a}\left(c_{\rm s1}^{2} -
2w_{\rm 1}\right) \nonumber\\
 - 3a\h\left(\rho_{\rm 1} + p_{\rm
1}\right)\frac{3\h^{2}\left(3c_{\rm s1}^{2} - 1\right) +
k^{2}\left(3c_{\rm s1}^{2} + 1\right) + 6\left(\h^{2} -
\h\dot{\h}a\right)}{k^{4} + \left(3\h^{2} +
k^{2}\right)\left(\h\dot{\h}a - \h^{2}\right)},\\
 B_{\rm 1} = - 3a\h\rho_{\rm 2}\left(1 + w_{\rm
1}\right)\frac{3\h^{2}\left(3c_{\rm s1}^{2} - 1\right) +
k^{2}\left(3c_{\rm s1}^{2} + 1\right) + 6\left(\h^{2} -
\h\dot{\h}a\right)}{k^{4} + \left(3\h^{2}
+ k^{2}\right)\left(\h\dot{\h}a - \h^{2}\right)},\\
 C_{\rm 1} = \frac{k^{2}c_{\rm s1}^{2}}{a^{2}} +
\frac{3}{a^{2}}\left(\dot{\h}\h a + 4\h^{2}\right)\left(c_{\rm
s1}^{2} - w_{\rm 1}\right) + 3\frac{\h^{2}}{a}\left(c_{\rm
s1}^{2}\right)\dot{} - 3\left(1 +
w_{1}\right)c_{\rm s1}^{2}\rho_{\rm 1} + \nonumber \\
 3\h^{2}\left(2 + 3c_{\rm s1}^{2}\right)\left(\rho_{\rm 1} +
p_{\rm 1}\right)\frac{k^{2} - 3\left(c_{\rm s1}^{2} - w_{\rm
1}\right)\left(3\h^{2} + k^{2}\right)}{k^{4} + 3\left(3\h^{2}
+ k^{2}\right)\left(\h\dot{\h}a - \h^{2}\right)} \nonumber\\
 -\left(k^{2} + 3\h^{2} - 6\h\dot{\h}a\right)\left(\rho_{\rm 1} +
p_{\rm 1}\right)\frac{k^{2} + 3\left[\h\dot{\h}a - \h^{2} -
3\h^{2}\left(c_{\rm s1}^{2} - w_{\rm 1}\right)\right]}{k^{4}
+ 3\left(3\h^{2} + k^{2}\right)\left(\h\dot{\h}a - \h^{2}\right)},\\
 D_{\rm 1} = -3\left(1 + w_{\rm 1}\right)c_{\rm s2}^{2}\rho_{\rm
2}+ \nonumber \\
 3\h^{2}\left(2 + 3c_{\rm s1}^{2}\right)\rho_{\rm 2}\left(1 +
w_{\rm 1}\right)\frac{k^{2} - 3\left(c_{\rm s2}^{2} - w_{\rm
2}\right)\left(3\h^{2} + k^{2}\right)}{k^{4} + 3\left(3\h^{2}
+ k^{2}\right)\left(\h\dot{\h}a - \h^{2}\right)} \nonumber\\
 -\left(k^{2} + 3\h^{2} - 6\h\dot{\h}a\right)\rho_{\rm 2}\left(1
+ w_{\rm 1}\right)\frac{k^{2} + 3\left[\h\dot{\h}a - \h^{2} -
3\h^{2}\left(c_{\rm s2}^{2} - w_{\rm 2}\right)\right]}{k^{4} +
3\left(3\h^{2} + k^{2}\right)\left(\h\dot{\h}a - \h^{2}\right)}.
\eea
The coefficients $A_{\rm 2}$, $B_{\rm 2}$, $C_{\rm 2}$ and $D_{\rm 2}$ have the same form, but with the interchange $1 \leftrightarrow 2$ in the subscripts.

\bibliography{osobliwoscizaburzenia}{}
\bibliographystyle{hieeetr}
\end{document}